# NEUTRON PHYSICS IN THE EARLY 1930S

Alberto G. De Gregorio[*]

> … I firmly hold him to see to it that he first makes sure as far as possible of the conclusion, through the senses, experiences, and observations, and afterwards searches for the means by which to prove it…
>
> GALILEO GALILEI (*Dialogue concerning the two chief world systems*: Salviati, the first day).

SEVEN DECADES HAVE elapsed since the artificial radioactivity was discovered. The outcome of producing new radioactive isotopes in laboratory represented a historic achievement in Physics, concerning both the foundations and the application of the latter: it caused the reappraisal of a sort of paradigm having consolidated for nearly four decades and leaving no chance for radioactive phenomena to depend on the outside, and it was the first step towards the human control of nuclear energy, sealed by the invention of the atomic pile and tragically marked by the fates of Hiroshima and Nagasaki.

In a posthumous treatise on experimental Physics, Antonio Genovesi, follower of the Enlightenment movement, stated:[1] «There exist two ways of dealing

---

[*] Research doctorate in Physics at the University «La Sapienza» of Rome (alberto.degregorio@roma1.infn.it). I'm grateful to Professor Fabio Sebastiani of «La Sapienza» for his advice in the preparation of the present work. This paper is extracted from my Ph.D. thesis in Physics, entitled «Dal "protone neutro" alla radioattività indotta: i neutroni a via Panisperna». The following abbreviations for journal titles are used: *CR*, Académie des Sciences, *Comptes rendus hebdomadaires des séances*; *JP*, *Journal de physique et le radium*; *PhR*, *Physical review*; *PRS*, Royal Society of London, *Proceedings*; *RiS*, *Ricerca scientifica*; *ZP*, *Zeitschrift für Physik*.

[1] Antonio Genovesi, *Elementa physicæ experimentalis usui tironum aptatæ* (Neapoli, 1779), auctoris præfatio, 5. «Verum cum duæ sint viæ & rationes res physicas pertractandi, una

with physics, the one historical in manner, and the other one by resorting to geometrical formulae and calculations. I have resolved on choosing the former, in order to make even the minds inexpert, and yet keen on natural things, imbued with the chief notions of the world, and predispose them to a more sublime knowledge». In a similar tenor Ettore Majorana, in the inaugural lecture to his course on theoretical physics held in 1938, warned:[2] «Two nearly opposite methods exist in expounding quantum mechanics as is at present. The one is the so called historical method: and it explains how, through precise and almost immediate indications arising from experience, the first idea of formalism appeared; and how the latter was subsequently developed, obeying to the quest of an internal necessity, rather than to the need of accounting for new and crucial experimental facts. The other one is the mathematical method, describing from the very beginning the quantum formalism in its most general and therefore most clear approach, and showing only afterwards the criteria for applying it. Each of these two methods, if used in an exclusive fashion, presents very serious inconveniences».

The present paper is devoted to the events related to the dawning neutron physics, and to Enrico Fermi's discovery and early investigation of radioactivity induced by neutrons. Following Majorana and Genovesi, that the most relevant subjects of Physics – even in the case that they concern experiments – demand a wide and deep analysis not solely devoted to a strictly technical and scientific outline, but

---

historicum in morem, altera geometricis formulis & calculis, priorem hic ego magis premere decrevi, quo rudes adhuc animos, utut naturæ rerum studiosos, primis Mundi notitiis imbuam, & ad sublimiorem intelligentiam præparem».

[2] Ettore Majorana, *Lezioni all'università di Napoli* (Napoli, 1987), 171. «Per esporre la meccanica quantistica nel suo stato attuale esistono due metodi pressoché opposti. L'uno è il cosiddetto metodo storico: ed esso spiega in qual modo, per indicazioni precise e quasi immediate dell'esperienza, sia sorta la prima idea di formalismo; e come questo si sia successivamente sviluppato in una maniera obbligata assai più dalla necessità interna che non dal tenere conto di nuovi decisivi fatti sperimentali. L'altro metodo è quello matematico, secondo il quale il formalismo quantistico viene presentato fin dall'inizio nella sua più generale e perciò più chiara impostazione, e solo successivamente se ne illustrano i criteri applicativi. Ciascuno di questi due metodi, se usato in maniera esclusiva, presenta inconvenienti molto gravi».

also regarding the historical perspective, a resort to the «historical method» rather than to the «mathematical method» will prevail here.

In fact, the very physicists who allowed such important phenomena to come to light strongly contributed also to their experimental deepening and theoretical understanding. At the same time, however, it is also true that so far the historical method has only partially clarified the circumstances in which the early discoveries concerning neutron-induced radioactivity ripened; within such a frame, indeed, further results appear to be still attainable, some historical aspects of dawning neutron physics deserving wider consideration than they have so far received.

It is to be hoped that, in the spirit of Majorana's and Genovesi's lesson, not only might historians benefit from recollection of a period having some forgotten aspects, both interesting and complex from the technical point of view, but also physicists themselves might benefit from facing strategies and solutions suggested and adopted at a time, when tools and interpretative categories differed so much from those in use in physics today.

In the present work a frame will be outlined, from which the discovery made by Fermi in March 1934 arose. The one who investigates the skies with a telescope, loses sight of the lines describing the constellations. Only by going back and looking at it with the naked eye, can he make out the lines joining the stars. In a not very dissimilar way, in this paper we are thoroughly investigating numerous works concerning neutron physics in the early 1930s, which might seem at first to have no common traits, except the fact of concerning neutron physics. Still, when we have reviewed the discovery of artificial radioactivity induced by neutrons, we will be in a position to make out the way in which they relate the one to the other, and to the discovery of neutron-induced radioactivity.

Many laboratories began to investigate neutron properties, straight after James Chadwick's experiments, paying attention particularly to the way in which neutrons interacted with matter; experiments were very soon performed, concerning nuclear reactions produced by neutrons, and neutron absorption and scattering. All the main laboratories then active in nuclear physics were reorganised after the discovery of the neutron. Researches flourished in a rapid succession in France, in England, and in Germany, starting from February 1932; the United States were apparently late in joining that line of research, but succeeded in bridging the gap soon and in making substantial contribution, leading to remarkable results.

Towards the mid of February 1932 Frédéric Joliot and Irène Curie observed polonium-beryllium sources generating some β traces in a Wilson chamber, and by means of a magnetic field of 1,300 gauss estimated the energy of those particles exceeding ten million electronvolts.[3] By resorting to the ionisation method (*i.e.* to an ionisation chamber) that they had already used in previous experiments, they noticed that the radiation from beryllium could project carbon and helium nuclei. Projection had been already observed to involve hydrogen,[4] and the Joliot-Curies concluded it was probably a general phenomenon, generated by high energy γ-rays coming from beryllium irradiated with α's. A paper soon followed, in which Chadwick drew attention to the projection phenomena involving light nuclei, and announced the possible existence of a neutron.[5] The Joliot-Curies promptly replied, and during the session of the *Académie des Sciences* of March 7 they stressed that Chadwick's hypothesis, according to which beryllium radiation consisted of particles of mass 1 and charge 0, should be reconciled in some way with «the emission of secondary electrons of high energy (from 2 to more than $10 \cdot 10^6$ eV) we have just pointed out».[6] During the same session Pierre Auger communicated that he had noticed the existence of electronic traces coming from the walls of a Wilson chamber, and generated by beryllium radiation. He stressed the difficulties in ascribing the projection of both high energy electrons and protons to the same kind of radiation: «Maybe the existence of two different types should be foreseen, the first one electromagnetic, producing the Compton effect, the other one consisting of those neutrons whose existence has recently been supposed by Chadwick, and which would project the positive nuclei».[7] Electronic traces associated with beryllium radiation will often capture physicists' attention later on.

---

[3] Irène Curie and Frédéric Joliot, "Effet d'absorption de rayons γ de très haute fréquence par projection de noyaux légers", *CR*, *194* (1932), 708-711.

[4] Irène Curie and Frédéric Joliot, "Émission de protons de grande vitesse par les substances hydrogénées sous l'influence des rayons γ trés pénétrants", *CR*, *194* (1932), 273-275.

[5] James Chadwick, "Possible existence of a neutron", *Nature*, *129* (1932), 312.

[6] Irène Curie and Frédéric Joliot, "Projections d'atomes par les rayons très pénétrants excités dans les noyaux légers", *CR*, *194* (1932), 876-877, on 877.

On March 15, 1932 Franco Rasetti, then in Berlin, sent to *Naturwissenschaften* a paper contradicting the hypotheses that beryllium radiation consisted solely of neutrons.[8] German laboratories relied on the work of scholars as eminent as Walter Wilhelm Bothe, Hans Geiger and Lise Meitner, and were in the van of neutron research; Berlin really was one of the places where experimental researches on the nucleus were most advanced in Europe. Anyway, we are focusing here on some other researches on neutrons, carried out in France and in England.

1. IN THE CURIES' SCHOOL

The *Institut du radium* could boast a decisive advantage over other institutes devoted to nuclear research, arising from the deep commitment lavished for tens of years by Pierre and Marie Curie: a huge amount of knowledge and experience, and the largest supply of polonium in the world, had been accumulated in Paris. Not only Irène Curie and her husband Joliot, but also physicists such as Auger, M. de Broglie, Leprince-Ringuet, F. Perrin, Thibaud, could benefit by so rich heritage, and the French school investigated thoroughly processes regulating the interaction of neutrons with matter. In particular, absorption and scattering by different elements were investigated, many of the important achieved results being communicated to the *Académie des Sciences*.

The experimental technique that was required was relatively simple, and consolidated rapidly. The basic arrangement consisted of an ionisation chamber or a cloud chamber, adjusted to detect neutrons by merely resorting to some paraffin: in fact, by just lining the chamber or filtering the radiation with paraffin, protons were ejected which revealed the presence of neutrons; at variance, the detection chamber was filled with gas containing a large amount of hydrogen.

---

[7] Pierre Auger, "Sur la projection de noyaux légers par les rayonnements ultra-pénétrants de radioactivité provoquée. Trajectoires photographiées par la méthode de Wilson", *CR*, *194* (1932), 877-879, on 879.

[8] Franco Rasetti, "Über die Natur der durchdringenden Berylliumstrahlung", *Naturwissenschaften*, *20* (1932), 252-253.

The sources were another fundamental component of the instrumentation then in use, obviously. The Joliot-Curies possessed supplies of polonium emitting α-particles with an activity of about one hundred millicurie, thus exceeding for one order of magnitude the activity of the sources used in England and in Germany. In a very short time beryllium powder started being irradiated in some cases with emanation of radium (*i.e.* radon) and its decay products, in place of polonium. Really, a strong γ-radiation from radon and its products, and alpha particles far from being monochromatic, sometimes rendered emanation-beryllium (Em + Be) neutron sources little or not in the least suited to neutron researches; notwithstanding this, radon-beryllium sources were so much simpler to obtain than polonium-beryllium ones, that the latter advantage very often made up for the former disadvantages.[9] Thus, just some ten days after Chadwick's paper appeared, Maurice de Broglie announced that he and other physicists had resorted to radon-beryllium sources.[10]

Two works resorting to radium emanation mixed up with beryllium powder were communicated to the *Académie des Sciences* on May 9, 1932. The first one was by Maurice de Broglie and Louis Leprince-Ringuet:[11] they placed a piece of lead as a screen between a Em + Be source and a ionisation chamber lined with paraffin, and investigated the scattering of neutrons by putting sideways, and then taking away, some blocks of different materials (lead, potassium chloride, paraffin) as side-screens. They observed a strong effect, noticing that while the scattering due to the lateral lead remained of the same magnitude as that due to the walls and to the objects in the laboratory, scattering by screens of paraffin and KCl produced a sharp increase in number of the deflections in the recording oscillograph. Moreover, de

---

[9] In addition, one should bear in mind that within some tens of minutes radon reaches equilibrium with two of its decay products which in turn emit alphas. Thus, one millicurie of radon can be considered equivalent to about three millicurie of polonium, relating to the number of produced neutrons. For what concerns half-lives, instead, that of polonium is 138 days long, that of radon 3.8.

[10] Maurice de Broglie, at the foot of: Auger (ref. 7), 879-880.

[11] Maurice de Broglie and Louis Leprince-Ringuet, "Sur la dispersion des neutrons du glucinium et l'existence de noyaux de recul provoqués par le lithium excité", *CR*, *194* (1932), 1616-1617.

Broglie and Leprince-Ringuet discovered that emanation of radium caused also lithium to emit neutrons.

The second work communicated on May 9 and involving a Em + Be source concerned another remarkable phenomenon, revealing how neutrons interact with matter. Jean Thibaud and F. Dupré la Tour carried out some experiments with a ionisation chamber lined with paraffin: «In this way one can establish the absorption coefficient for neutrons alone, independently of the effects due to the less penetrating radiation (γ-rays) that can be simultaneously emitted by beryllium».[12] Beryllium radiation proved to be remarkably penetrating – «more than one tenth of the incident radiation still persists across a shield of 30 cm of lead» – and to have a heterogeneous absorption in lead: across the ten centimetres between 2.5 and 12.5 cm of thickness, the number of neutrons was reduced by three quarters, whilst it was reduced by solely one half between 12 and 22 cm of lead. The same phenomenon was noticed with other materials as well. Thibaud and Dupré la Tour's account was based on a different behaviour of neutrons, depending on their velocity: less swift neutrons would suffer an angular dispersion much stronger than fast ones did, thus being much more absorbed. It was not a novelty that neutrons were in fact heterogeneous, since the Joliot-Curies had already pointed up a continuous spectrum for them.[13] Nevertheless, it was the first time, as far as I know, that somebody guessed a different behaviour for neutrons, depending on their velocity; it would certainly not be the last one. On June 20 the Joliot-Curies, together with P. Savel, communicated a new result about neutrons from lithium. That is, if neutrons were ejected by the α-particles coming from radium emanation, they were able to penetrate through many centimetres of lead; rather, neutrons ejected by the α's from polonium were fully absorbed by 5 mm of lead, and could only project protons not swift enough to be revealed: «Thus, radiation excited by polonium α-rays certainly consists of neutrons of feeble velocity and of feeble penetration».[14] Some months later, the Joliot-Curies would emphasise that the results communicated by M. de Broglie and Leprince-Ringuet on May 9, «compared with ours, are a clear example of the increase of the penetration of neutrons with the energy of the incident α-rays».[15]

---

[12] Jean Thibaud and F. Dupré la Tour, "Sur le puvoir de pénétration du rayonnement (neutrons) excité dans le glucinium par les rayons α", *CR*, *194* (1932), 1647-1649, on 1647.

[13] Irène Curie and Frédéric Joliot, "Sur la nature du rayonnement pénétrant excité dans les noyaux légers par les particules α", *CR*, *194* (1932), 1229-1232.

On July 11, de Broglie and Leprince-Ringuet went through absorption results again, reporting that neutrons, if emitted from an emanation-boron source, passed through lead much more easily than through paraffin. So much, that they referred to a «transparency of lead».[16]

Really remarkable is also a work by Pierre Auger, communicated the following week.[17] Auger resorted to neutron sources of two different kinds: polonium-beryllium, or else radon-beryllium; as a measuring instrument, a Wilson chamber filled with hydrogen saturated with water vapour. He observed very short tracks due to recoil protons with energies between 30 and 500 keV, and forming small angles to the direction of the incoming neutrons; such neutrons, therefore, could not have energies exceeding those values too much. In addition, traces were present many tens of centimetres of length, revealing the presence of fast protons too. Auger did not observe any traces of length intermediate between those of the two previous groups of protons: in other words, he could establish the existence of a definite and clear-cut group of «slow neutrons».

In order to investigate neutron scattering, Auger carried out two sets of measurements. In a first arrangement, a metallic screen some centimetres thick was placed between the neutron source and the cloud chamber, and a bulk of 100 kg of copper encompassed them almost completely; in the second case the copper shell and the median screen were removed. With the copper shell in place, and notwithstanding the simultaneous presence of the thick median screen, the neutron tracks – particularly those shorter than 20 mm (corresponding to energies less than 300 keV) – increased in number, becoming some two or three times more frequent than with the sole radioactive source in the vicinity: Auger deduced an effect of «gathering» (*concentration*) of neutrons inside the cavity. At the end of his memoirs, Auger also contemplated that in some cases the energy released in the disintegration of beryllium might be shared between the ejected neutron and a γ-ray.

---

[14] Irène Curie, Frédéric Joliot and P. Savel, "Quelques expériences sur les rayonnements excités par les rayons α dans les corps légers", *CR*, *194* (1932), 2208-2211, on 2210.

[15] Irène Curie and Frédéric Joliot "Preuves expérimentales de l'existence du neutron", *JP*, *4* (1932), 21-33, on 30.

[16] Maurice de Broglie and Louis Leprince-Ringuet, "Sur les neutrons du bore excité par l'émanation du radium", *CR*, *195* (1932), 88-89, on 89.

[17] Pierre Auger, "Émission de neutrons lents dans la radioactivité provoquée du glucinium", *CR*, *195* (1932), 234-236.

Auger came to more complete results on neutron scattering six months later.[18] His report communicated to the *Académie des Sciences* on 16 January, 1933 gains great significance with regard to Fermi's experiments to come.[19] Auger again reported on the considerable proportion of short proton tracks (about 30%). Note that Auger surrounded the cloud chamber and neutron source with a 'shield' of paraffin 5 to 10 cm thick, in order to get rid of the effects due to the environment. Short proton tracks were much reduced in number with the paraffin, far more than long ones; nevertheless, by merely placing about a 1 dm³ of copper scatterer inside the shield of paraffin, the short tracks in the Wilson chamber returned almost as numerous as before. Auger observed a similar effect also with iron, aluminium or lead in place of copper, thus concluding that the group of less swift neutrons might arise from γ emission in the disintegration of beryllium, as well as from the loss of energy of fast neutrons scattered by matter. The trouble was that a heavy nucleus cannot sensibly reduce neutron velocity by one single elastic collision. Thus Auger concluded that neutrons should lose their energy on account of inelastic nuclear collisions.

As the Joliot-Curies already did, Auger noticed some tracks belonging to swift β-particles, some of which were apparently directed towards the neutron source.

---

[18] Pierre Auger, "Sur la diffusion des neutrons. Chocs non élastiques sur les noyaux", *CR*, *196* (1933), 170-172.

[19] On October 1934 Fermi and his group became concerned for the effects of the environment on the radioactivity induced by neutrons. While investigating such effects,

According to Auger, such β-rays would be produced by the γ-rays emitted, in a sort of «nuclear fluorescence», by the environmental matter hit by swift neutrons.[20]

Further experiments revealed the phenomena connected with the passage of beryllium radiation through matter to be more complex than what appears from the results reported so far. Before deepening such topics, we shall widen the frame concerning early neutron physics, dealing with the researches carried out in England, in Rutherford and Chadwick's school.

2. THE INTERACTION WITH NUCLEI

Really, the artificial excitation of γ-rays from nuclei was not a novelty, John C. Slater having obtained a weak γ-radiation from tin and lead by means of α-particles from radium emanation in 1921.[21] Moreover, such γ-emission occurred quite regularly in the artificial transmutation of the light elements; not only did it occur in those transmutation processes produced by α-particles, actually, but also in those which, as was soon realised, were produced by neutrons. For example, γ-rays were emitted in the reaction $n^1 + N^{14} \rightarrow B^{11} + He^4$, discovered by Norman Feather in February 1932.[22]

Disintegration produced by neutrons deeply differed from that produced by α-particles; nevertheless, some important features were common to both. We are dealing now with both phenomena, exploiting those analogies which might help to catch some of the features which were ruling the interaction of neutrons with nuclei, particularly those features which would be involved in radioactivity induced by neutrons.

At the head of all the nuclear transmutation processes stood the reaction

---

  Fermi resorted to a screen of paraffin, thus readily discovering the great efficiency of slow neutrons in inducing radioactivity.

[20] See also: Pierre Auger and Gabriel Monod-Herzen, "Sur l'émission de neutrons par l'aluminium sous l'action des particules α", *CR*, *196* (1933), 543; Curie and Joliot (ref. 13), 1231.

[21] John C. Slater, "The excitation of γ radiation by α particles from radium emanation", *Philosophical magazine*, *42* (1921), 904-923.

[22] Norman Feather, "The collisions of neutrons with nitrogen nuclei", *PRS*, *136* (1932), 709-727.

$$He^4 + N^{14} \rightarrow O^{17} + H^1 \qquad [1]$$

discovered by Ernest Rutherford in April 1919, when he observed that α-particles from radium C ($Bi^{214}$) ejected some hydrogen nuclei from nitrogen, thus producing scintillation on a zinc sulphide screen.[23] In a following letter to *Nature* Rutherford and Chadwick announced that also boron, fluorine, sodium, aluminium and phosphorus ejected protons, but no effect was noticed with elements whose atomic number was given by 4*n*, where *n* was a whole number.[24]

At the end of 1924 Patrick M.S. Blackett supplied the 'photographic proof' of the reaction [1], by taking photographs of the tracks of the impinging α-particles – emitted by thorium B + C ($Pb^{212}$ and $Bi^{212}$) – and of the residual nuclei, in an automatic Wilson chamber filled with nitrogen diluted with oxygen.[25] Blackett also proved that there was no trace of α-particles escaping from the nucleus after having ejected the proton: effective alphas were captured, really. In 1927 Bothe and Hans Fränz investigated proton emission by resorting to a point Geiger counter,[26] by means of which the following year they noticed the existence of two distinct groups of protons from boron.[27]

The topic of the energy distribution of the disintegration protons was tackled by Chadwick and George Gamow from a theoretical point of view also, and related to different modes of interaction of nuclei with incident α-particles.[28] The four wave functions of the α-particle and of the proton, before and after the disintegration of the nucleus, entered an integral leading to the disintegration probability. In case of capture (possibly resonant) of α-particles, protons would have a line spectrum, being distributed nearly uniformly in all directions; if α's were not captured, protons would not be emitted uniformly in all directions, and would have a continuous spectrum.

---

[23] Ernest Rutherford, "Collision of α particles with light atoms. *IV*. An anomalous effect in nitrogen", *Philosophical magazine*, *37* (1919), 581-587.

[24] Ernest Rutherford and James Chadwick, "The disintegration of elements by α-particles", *Nature*, *107* (1921), 41.

[25] Patrick M.S. Blackett, "Ejection of protons from nitrogen nuclei, photographed by the Wilson method", *PRS*, *107* (1925), 349-360.

[26] Walter Wilhelm Bothe and Hans Fränz, "Atomzentrümmerung durch α-Strahlen von Polonium", *ZP*, *43* (1927), 456-65.

[27] Walter Wilhelm Bothe and Hans Fränz "Atomtrümmer, reflektierte α-Teilchen und durch α-Strahlen erregte Röntgenstrahlen", *ZP*, *49* (1928), 1-26.

[28] James Chadwick and George Gamow, "Artificial disintegration by α-particles", *Nature*, *126* (1930), 54-55.

The emission of discrete groups of protons in the disintegration of light elements proved to be quite a general phenomenon: by a small ionisation chamber, one or more groups were observed not only from boron, but also from aluminium and other elements, such as nitrogen, fluorine and phosphorus; protons from sodium, rather, showed a heterogeneous distribution.[29]

A detailed account of such processes was provided in Gamow's report, read in the nuclear conference held in Rome in October 1931.[30] After having extensively reported on nuclear theory, Gamow – who only sent a written account, read by Max Delbruck during the conference – dealt with some particular cases of interaction with α-particles, namely nuclear excitation and nuclear disintegration occurring with or without capture. Some difficulties arose in case of capture of the incident α-particle, with radiation of energy: as Gamow stressed, in fact, the probability of such process to occur roughly equalled the ratio of the time needed by the α-particle to cross the nucleus, to the average period of γ-emission. Now, the former interval was given by the nuclear radius divided by the α-particle velocity (thus being of the order of $10^{-21}$ s), whereas the latter was of the order of $10^{-16}$ s. The probability for radiative capture to occur resulted to be only 0.001% per penetration of the potential barrier. Gamow concluded that the conditions were most favourable for radiative capture in case of resonance, «as in this case the α-particle will stay longer time inside the nucleus oscillating many times between the walls».[31] The inconvenience in this case was that the resonance region would be very narrow, so that Gamow ended up by guessing that the artificial radiation observed from lithium and beryllium resulted from an excitation process occurring without capture.

---

[29] James Chadwick, J.E.R. Constable and E.C. Pollard, "Artificial disintegration by α-particles", *PRS*, *130* (1931), 463-489.

[30] George Gamow, "Quantum theory of nuclear structure", *Convegno di fisica nucleare (ottobre 1931)*, (Roma, 1932), 65-81.

The discovery of the neutron considerably widened the subject of the experimental enquiries we have so far dealt with: on the one hand, the survey of $\alpha$-induced reactions spread out, since not only protons, but neutrons as well turned out to be ejected from nuclei by alphas; on the other hand, neutrons too turned out to be good for producing nuclear disintegration, so that alphas ceased all at once to be the sole disintegration agents.

Feather had obtained[32] the first neutron-induced nuclear reaction

$$n^1 + N^{14} \rightarrow B^{11} + He^4 \qquad [2]$$

the day before Chadwick sent to *Nature* the letter announcing the discovery of the neutron. Feather was one of Chadwick's collaborators in Cambridge and, resorting to a Wilson chamber, he had been searching for «certain cases of artificial disintegration»[33] for six months, since mid-1931. On February 16, 1932, by employing a Po + Be source and an expansion chamber filled with nitrogen diluted with about 4% by volume of oxygen, «an entirely new phenomenon was observed, namely, examples of paired tracks having a common origin».[34] Feather, in fact, had taken the first photographs in which two distinct tracks sprang ahead from a single point in the gas, no track from the source. In that same paper he stated that those photographs had been interpreted, a week after February 16, as the signs of the disintegration of nitrogen nuclei by the incident radiation. The first public announcement of Feather's results came on March 18, 1932, as Rutherford communicated them before the Royal Institution.[35] Feather's paper, instead, was received by the Royal Society on May 10. According to Feather, the disintegration produced by the beryllium radiation provided «further support for the hypothesis of its neutron (particular) nature».[36]

---

[31] Ibid., 80.

[32] Feather (ref. 22).

[33] Ibid., 710.

[34] Ibid.

[35] Ernest Rutherford, "Origin of the gamma rays", *Nature*, 129 (1932), 457-458.

It is really noteworthy that Chadwick wrote his renowned letter to *Nature*, announcing the possible existence of a neutron, on February 17, 1932, just the day after that, in his own laboratories, one of his collaborators had taken the first photographs giving evidence of disintegration produced by beryllium radiation. Before the beginning of 1932 the only known nuclear disintegration was that produced by material particles, such as $\alpha$-particles. It is really plausible, then, that Feather's experiments strongly supported Chadwick in interpreting the penetrating radiation from beryllium as to consist of neutrons, thus playing a crucial role in the discovery of the neutron. Nevertheless, historiography does not seem to have acknowledged much regard to such circumstance.

Disintegration processes produced by neutrons proved to be rather complex: even though «detailed analysis […] shows that when disintegration occurs with capture of the neutron an $\alpha$-particle is expelled»[37] in accordance with [2], nevertheless there was still room for the reactions $n^1 + N^{14} \rightarrow C^{14} + H^1$ and $n^1 + N^{14} \rightarrow C^{13} + H^2$ to take place; moreover, the possibility that disintegration without capture of the neutron might also take place was not excluded either.[38] Feather also mentioned the possibility of «a more drastic disintegration process» to take place; it would consist in the breaking up of the nitrogen nucleus, ending up with the production of two nuclei of almost the same mass: $Be^8$ and $Li^7$ in case of neutron capture, or $Be^8$ and $Li^6$ if without capture. Still, Feather did not deepen that topic, since «it would be unprofitable to pursue further such a suggestion as this».[39] Some years later, similar comments would brand Ida Tacke Noddack and her proposal of a drastic disintegration of the uranium nucleus.

---

[36] Feather (ref. 22).

[37] Ibid.

[38] At about the end of 1933 Feather found that at least the second of such reactions – that resulting in the ejection of heavy hydrogen – really took place (Norman Feather, "Collisions of neutrons with light nuclei. Part II", *PRS*, *142* (1933), 689-709, on 708). Harkins, Gans and Newson, Meitner and Philipp, and Kurie would obtain some nitrogen disintegration by neutrons.

[39] Feather (ref. 22), 722.

As in the case of experiments carried out with α-particles, in experiments carried out with neutrons γ-rays were produced as well. Thus, γ-emission turned out to be a general feature of nuclear transmutation processes. Electronic tracks were also produced in the Wilson chamber, but much care was used in Cambridge in dealing with high energy electronic tracks related to the passage of neutrons.[40]

In a letter to *Nature*, Feather dealt with the disintegration of oxygen he had observed in a Wilson chamber, taking place with neutron capture and α-emission, in accordance with the reaction:

$$n^1 + O^{16} \rightarrow C^{13} + He^4 \qquad [3]$$

Generally, nuclei so produced were in an excited state.[41]

To sum up, both experiments concerning α-particles and experiments concerning neutrons shared some common features: it had been proved that both alphas and neutrons could disintegrate light elements; moreover, the hypotheses that the nucleus, in most cases, captured the incident particle and ejected another corpuscle was accepted. Such ejection often occurred with emission of γ-rays, even though, in the case that the α's caused the ejection of neutrons, tracks due to fast electrons were also produced, which had been still not fully accounted for.

---

[40] Feather overlooked high energy β traces in the Wilson chamber, by simply deferring the question to a work by P.I. Dee (another Chadwick's collaborator). In turn, Dee studied only short electronic tracks related to the passage of neutrons, establishing solely that fast β's could cause difficulties, since they themselves produced short secondary tracks (P.I. Dee, "Attempts to detect the interaction of neutrons with electrons", *PRS*, *136* (1932), 727-737).

[41] Norman Feather, "Artificial disintegration by neutrons", *Nature*, *130* (1932), 237. Feather had intended to investigate the action of neutrons also on fluorine. Nevertheless, *de facto* he only searched for tracks in a Wilson chamber (filled with carbon tetrafluoride) which would have revealed the ejection of neutrons by fluorine irradiated by alphas (Norman Feather, "Collisions of α-particles with fluorine nuclei", *PRS*, *141* (1933), 194-209). Feather was spurred to such investigation by some experiments carried out by Chadwick: even if the latter, by resorting to a polonium source and to an ionisation chamber, had apparently discovered that neutrons could be ejected from fluorine (and from magnesium too), still, his results were doubtful. Afterwards, in February 1933, the Joliot-Curies would show by means of a ionisation chamber that fluorine (like aluminium as well) ejected neutrons (James Chadwick, "The existence of a neutron", *PRS*, *136* (1932), 692-708; Irène Curie and Frédéric Joliot, "Sur les conditions d'émission des neutrons par action des particules α sur les éléments légers", *CR*, *196* (1933), 397-399.

On May 25, 1933, Chadwick held his *Bakerian lecture*, devoted to *The neutron*. He collected there many results physicists had attained on neutrons in almost one year and a half, mainly mentioning those results obtained in Cambridge.[42] Chadwick stated that «the most obvious properties of the neutron are its ability to set in motion the atoms of matter through which it passes and its great penetrating power».[43] In addition to these «obvious properties», among the topics dealt with by Chadwick we may recall: the higher efficiency of radon-beryllium neutron sources with respect to polonium-beryllium ones; a list enumerating the elements which ejected neutrons;[44] the empirical rule $A \geq 2Z$ which, according to Chadwick, prevented $He_2^4$, $C_6^{12}$, $N_7^{14}$, $O_8^{16}$, $B_5^{10}$, $Ne_{10}^{20}$, and $Mg_{12}^{24}$ from capturing one α-particle and ejecting one neutron.[45] Chadwick, moreover, dealt with the very interesting cases of fluorine and aluminium.

Still, we are mainly concerned here for a calculation reported by Chadwick in his *Bakerian lecture*, rather than for the previously mentioned topics and further topics also dealt with there by Chadwick (some of which we are examining closely in the next paragraph). Such calculation had already been carried out by Harrie

---

[42] James Chadwick, "Bakerian lecture. The neutron", *PRS*, *142* (1933), 1-25.

[43] Ibid., 1.

[44] Such list comprised: lithium, beryllium, boron, fluorine, neon, sodium, magnesium, and aluminium.

[45] Similar reasoning had been already published by Georges Fournier in 1930, and considered again by F. Perrin two years later (Georges Fournier, "Sur une classification nucléaire des atomes en relation avec leur genèse possible et leur désintégration radioactive", *JP*, *1* (1930), 194-205; Francis Perrin, "L'existence des neutrons et la constitution des noyaux atomiques légers", *CR*, *194* (1930), pp. 1343-1346).

Massey,[46] and gave the neutron-proton scattering cross-section as a function of neutron velocity. Attaining detailed knowledge on neutron-proton scattering as a function of the velocity would strongly help, in obtaining information on such a still unsolved problem, as the nature of the neutron was. From the experimental point of view, investigation into the less swift neutrons had already revealed that «the radius for the proton collisions continues to increase as the velocity of the neutron decreases».[47] Now, the calculation of the neutron-proton elastic cross section led to the formula:

$$Q \approx \frac{h^2}{\pi M^2 v^2} \qquad [4]$$

where M is the reduced mass, and v the relative velocity; thus, also from the theoretical point of view the neutron-proton elastic cross section increased as the neutron velocity decreased.

Chadwick pointed out that, besides elastic collisions, inelastic collisions also took place. He reported Feather's results on the transmutation produced by neutrons in nitrogen, in oxygen, and in carbon, and stressed that γ-rays were often emitted in such disintegration processes.

The last paragraph was devoted to the positive electrons: Carl D. Anderson[48] had discovered the positron some months earlier, while investigating cosmic radiation, but now Chadwick pointed out that beryllium radiation as well apparently produced positive electrons, in its passage through matter. Nevertheless, Chadwick stressed that, if that was really the case, then it could well be that not only γ-rays, but neutrons too produced positives. We are dealing with positive and negative electrons related to the beryllium penetrating radiation in the next paragraph.

3. SYMMETRIES

---

[46] Harrie S.W. Massey, *"The passage of neutrons through matter"*, *PRS*, *138* (1932), 460-469.

[47] Chadwick (ref. 42), 18.

[48] Carl D.Anderson, "The apparent existence of easily deflectable positives", *Science*, *76* (1932), 238-239.

On February 22, 1932, when the Joliot-Curies communicated that they observed high energy β-rays in a Wilson chamber placed in a magnetic field,[49] penetrating radiation from beryllium was generally considered as to comprise only photons, with energies up to five tens of MeV – in fact, it was before Chadwick's letter appearing on *Nature*. Thus, it was obvious for the Joliot-Curies to ascribe the electronic tracks to the Compton effect. On march 7, after the appearance of Chadwick's letter, the Joliot-Curies, as well as Auger, pointed out that such β-traces suggested that beryllium radiation did not only consist of neutrons.[50] To tell the truth, if neutrons could not account for such electronic tracks, still – standing the knowledge of the beginning of 1932 – gamma-rays of many MeV as well could not give a full account for tracks with such very peculiar properties. The peculiarities of the electronic tracks generated by the beryllium radiation were described in an explicit way only on the following April 11:[51]

> Many tracks resemble electronic tracks, though bent in the opposite direction: it is very probable that they belong to electrons emitted in the direction opposite to that of the incident beam, having sometimes a very high energy, $2 \cdot 10^6$ eV. Such rays can't be electrons hit by the primary photons.

Irène Curie and Frédéric Joliot ascribed the production of what they interpreted as backwards electrons to secondary rays. Moreover, they showed that a screen of lead placed on the path of the penetrating radiation reduced the number of electron tracks in the chamber, thus coming to the conclusion that electrons should be produced by γ-rays; still, the Joliot-Curies did not clearly state if the number of 'retrograde' tracks too was reduced in proportion, and then if also 'backwards' electrons were produced by γ-rays.

    Indeed, in November the Joliot-Curies stated that such high energy electrons apparently heading for the neutron source should be ascribed to the neutrons. In fact, according to them:[52]

---

[49] Curie and Joliot (ref. 3).

[50] Curie and Joliot (ref. 6); Auger (ref. 7).

[51] Curie and Joliot (ref. 13), 1230.

[52] Curie and Joliot (ref. 15), 31.

> It is scarcely probable that they consist of Compton electrons, projected by scattered γ-rays, otherwise one would be led to ascribe very high quantum energy to such a radiation. One might guess that some of the electrons projected forward have been reflected by the walls of the apparatus and come backwards. Nevertheless, stereoscopic analysis apparently points out that some of them sprang up from the gas.

Curie and Joliot's explanation for the appearance of such electrons did not substantially differ from the account that, as we have already seen, Auger would suggest in January 1933:[53] in fact, the latter would consider inelastic collisions of neutrons exciting the nuclei and the subsequent radiation of γ-rays producing Compton electrons, also backwards; the Joliot-Curies, instead, considered Compton electrons produced by the γ-rays emitted in neutron-induced transmutation processes, such as those observed with nitrogen by Feather.

To sum up, the penetrating radiation from beryllium generated electronic tracks in the cloud chamber. At first, they were ascribed to the γ-rays coming from the beryllium, and producing Compton effect, but by applying a magnetic field some tracks showed an inverse bending. That forced people to revise the previous account, in particular for what concerned the provenance of γ-rays which were supposed to produce the Compton electrons: γ's would not come all from the disintegration of the beryllium, but some should follow on transmutation or else upon the deexcitation processes, produced in nuclei by the neutrons themselves. At this stage, a new event occurred: the experiments carried out by Anderson, and those by Blackett and Giuseppe Occhialini, showed that cosmic radiation could produce positive electrons, when it passed through matter.[54]

The link that, since the Rome conference in 1931, had connected the two penetrating radiation – that of cosmic origin and that from beryllium – became topical again in 1933, when Chadwick, Blackett and Occhialini obtained new evidence for the positive electron by proving that beryllium radiation too, when passing through a lead screen, could produce positrons.[55] They placed a polonium-beryllium source close to the wall of a cloud chamber, on the inside of which was fixed a target of lead 2 mm thick. A magnetic field was applied, of about 800 gauss.

---

[53] Auger (ref. 18).

[54] Anderson (ref. 48); Patrick M.S. Blackett and Giuseppe P.S. Occhialini, "Some photographs of the tracks of penetrating radiation", *PRS*, *139* (1933), 699-726.

> Most of the tracks […] were, from the sense of their curvature, clearly due to negative electrons, but many examples were found of tracks which had one end in or near the lead target and showed a curvature in the opposite sense. Either these were due to particles carrying positive charge or they were due to negative electrons ejected in remote parts of the chamber and bent by the magnetic field so as to end on the lead target. Statistical examination of the results supports the view that the tracks began in the target and therefore carried a positive charge.

Chadwick, Blackett and Occhialini wondered in which manner the positive electrons could be produced, and also «whether they arise from the action of the neutron emitted by the beryllium or from the action of the accompanying $\gamma$-radiation».

This work threw a new light upon the 'retrograde' tracks produced by beryllium radiation, since such tracks could well be produced by positive electrons going forward, instead of being just due to 'retrograde' negative electrons. Notwithstanding this, the letter to *Nature* did not set everything clear. First of all, it did not clearly establish which of the two from beryllium, if neutrons or photons, were producing the positive electrons; moreover, it reported tracks having one end in the lead screen, or else *near* it: *i.e.*, there was a statistical support, but no definite proof that such tracks really all belonged to forward positive electrons.

Chadwick, Blackett and Occhialini's discovery spurred the Joliot-Curies on to re-examine the photographs they had taken the year before. Regrettably, at the time the source had been placed too distant from the cloud chamber, to let them now decide with respect to retrograde tracks. Thus, they decided to repeat their experiments.[56]

---

[55] James Chadwick, Patrick Blacket, and Giuseppe Occhialini, "New evidence for the positive electron", *Nature*, *131* (1933), 473.

[56] Irène Curie and Frédéric Joliot, "Contribution à l'étude des électrons positifs", *CR*, *196* (1933), 1105-1107; Irène Curie and Frédéric Joliot, "Électrons de matérialisation et de transmutation", *JP*, *4* (1933), 494-500.

A hole was made in the glass walls of the Wilson chamber, and then the hole was closed by an aluminium leaf. A thin layer of lead was housed in the inside. A source of neutrons and photons, consisting of beryllium irradiated with 100 mCi of polonium, was placed close to the chamber, and it caused tracks belonging to both positive and negative electrons to come out from the lead. A magnetic field of 1,100 gauss was then applied: every 10 negative electrons from lead, 2.8 positive electrons also appeared from lead; 1.76 electrons, positive as well as negative, came instead from the glass walls of the chamber. By replacing the lead layer with an aluminium one, 2 mm thick, the number of positive electrons passing through the layer decreased, whereas the number of electrons from the glass scarcely varied: every 10 negative electrons, there appeared only 0.53 positive electrons; 1.3 electrons instead, both positive and negative, came from the glass walls. The Joliot-Curies came to the conclusion that the positive electrons whose trajectories passed through the lead layer were produced in the very lead, actually.

The Joliot-Curies also proved that the positive electrons were produced by $\gamma$ rays, and not by neutrons: a screen of lead 2 cm thick was enough to reduce by 50% the electrons which were produced in the lead layer placed inside the chamber; moreover, also the $\gamma$-rays from thorium C" ($Tl^{208}$) could produce positive electrons in the lead layer. The Joliot-Curies called such electrons produced by high energy photons passing through matter «materialisation electrons» (*élèctrons de matérialisation*).

Two remarks should be made on the Joliot-Curies' experiments just described. The first remark is that the radiation from beryllium could well produce positive electrons in heavy elements like lead, but such property is not sufficient for «positive or negative» electrons springing from the glass walls of the chamber to be accounted for: the Joliot-Curies did not state at all whether also those tracks were – and to what extent – reduced in number, when the lead screen 2 cm thick was added. The second remark is that when the aluminium layer 2 mm thick was placed close to the neutron source, it may well be, *a posteriori*, that the aluminium was activated by the neutrons. In fact, the Joliot-Curies did not report absolute numbers; instead, they always referred the number of the investigated electrons to ten negative electrons coming from the layer. Still, in this way they provided no proof that the relative number of positive electrons passing from 2.8 to 0.53, with the aluminium layer in place of the lead one, signified a real reduction of the positive electrons: in fact, such reduction could be ascribed, in part at least, to an *increase* in number of the negative electrons when an aluminium layer was used. In other words, the tracks coming from the glass walls threw some 'shadows' on the electronic tracks related to beryllium radiation; what's more, signs of artificial radioactivity induced by neutrons might possibly become visible as the Joliot-Curies were carrying out those experiments.

The Joliot-Curies investigated not only materialisation electrons, but also «transmutation electrons», related to the transmutation processes produced by $\alpha$-particles in the light elements. We shall now go back to a couple of months before Chadwick, Blackett, and Occhialini's discovery, in order to get acquainted with transmutation electrons, thus to be able to thoroughly appreciate the role that those electrons would play in the experiments on artificial radioactivity to come.

On February 6, 1933, the Joliot-Curies showed that calcium fluoride and aluminium emitted a penetrating radiation consisting, in part at least, of neutrons.[57] They resorted to a ionisation chamber, filled with methane and connected to a Hoffmann electrometer, and observed that the radiation emitted by fluorite and aluminium irradiated with polonium $\alpha$-particles was much more absorbed by paraffin than by lead. That proved that such radiation certainly included neutrons, and the Joliot-Curies themselves acknowledged the importance of this fact:[58]

> It is known that $Al^{27}$ and $F^{19}$ can eject protons in transmutation processes produced by polonium $\alpha$-particles. *Thus, such nuclei can suffer transmutation of two different kinds, the one with proton emission, the other one with neutron emission (unless proton and neutron be emitted together)*

In nature, fluorine and aluminium have only one isotope each. In this respect, they differed from boron – which was already known to emit both protons and neutrons –, since in that case protons could be ascribed to $B^{10}$ while neutrons to $B^{11}$.

---

[57] Curie and Joliot (ref. 41). In those same days Feather published a paper reporting that his search, with a Wilson chamber, for neutrons emitted from fluorine had proved to be vain; Feather, "Collisions of $\alpha$-particles" (ref. 41).

[58] Curie and Joliot (ref. 41), 399.

Neutron emission from aluminium was confirmed by Auger and Monod-Herzen two weeks later, by resorting to a Wilson chamber lined with paraffin, and to a bulb containing aluminium powder and 280 mCi of radon emanation.[59] Furthermore, they proved that two groups of neutrons from aluminium also existed, besides the two groups of protons already known.

Next, the Joliot-Curies showed that 150 mCi of polonium caused neutron emission also from sodium and magnesium.[60] Fluorine and aluminium, and now sodium, all three were isotopically sheer elements, all three had a mass expressed by the formula $4n + 3$, and, according to the model then in use, differed the one from the others just by the addition of one $\alpha$-particle. Moreover, all three could eject both protons and neutrons. Fluorine could undergo the transmutation $\alpha + F^{19} \rightarrow Na^{22} + n$ into the new isotope $Na^{22}$; such isotope could be supposed unstable, so as to decay into $Ne^{22}$ through the capture of an «extranuclear electron» (*électron extranucléaire*). In the same way, sodium and aluminium too could transmute themselves into the still unknown $Al^{26}$ and $P^{30}$, respectively.

In the light of this double disintegration channel for elements like fluorine, sodium, and aluminium, the discovery of positively charged transmutation electrons, communicated[61] to the *Académie des Sciences* on June 19, 1933, becomes full of fascinating implications. The Joliot-Curies covered a 20 mCi polonium supply with a thin aluminium layer, and put it close to an orifice in the Wilson chamber placed in a magnetic field: positive electrons rose from aluminium. They got the same effect with boron, while, when they resorted to silver, paraffin or lithium in place of aluminium, positive electrons disappeared. Furthermore, the Joliot-Curies proved that the protons ejected from aluminium had no effect on the emission of positive electrons. In fact, one could absorb those protons by resorting, indifferently, to a silver or to an aluminium screen, and the relative number of positive electrons would not vary.

---

[59] Auger and Monod-Herzen (ref. 20).

[60] Irène Curie and Frédéric Joliot, "Nouvelles recherches sur l'émission des neutrons", *JP*, *4* (1933), 278-286.

[61] Irène Curie and Frédéric Joliot, "Électrons positifs de transmutation", *CR*, *196* (1933), 1885-1887.

Recalling that $Al^{27}$ and $B^{10}$ were known to undergo transmutation processes producing protons, the Joliot-Curies then reasoned that «sometimes the transmutation will take place with the emission of one neutron and one positive electron, in place of one proton»; thus, in the case of boron, one would no longer expect neutrons to come only from the heavy isotope $B^{11}$. In the following January, the Joliot-Curies discovered artificial radioactivity by observing that the emission of the positive electrons lasted beyond the $\alpha$-irradiation.

The Joliot-Curies drove to the extreme consequences the symmetry introduced by the discovery of the positron: they recovered a suggestion by Anderson[62] himself dating back to February, 1933 and stated that one should consider the proton – not the neutron anymore – as a compound particle, comprising one neutron and one positive electron. In this way they were faced with the $\beta$-decay to be accounted for, since there would be positive electrons in the nucleus, but no negative electrons. Still, the resulting nuclear model settled a thorny question related to the current nuclear model, which regarded the electron and proton as having a binding energy of the order of $mc^2$ (where $m$ is the electron mass) and then the mass of neutron as being smaller than that of the proton: considering the values then available for the nuclear masses, the beryllium nucleus, consisting of two $\alpha$'s and one neutron, would have been unstable. An 'elementary' neutron with a greater mass than the compound 'proton', instead, would have warranted the stability of the beryllium nucleus.[63]

---

[62] Carl D. Anderson, "The positive electron", *PhR*, *43* (1933), 491-494, on 494.

[63] Choosing the unit of nuclear masses so as to have He = 4, the mass of the beryllium nucleus turned out to be 9.011. According to the compound neutron model, the neutron mass should have been between 1.005 and 1.008, so that the beryllium nucleus would have had a mass greater than the sum of the masses of the three particles comprising it. An 'elementary' neutron of mass 1.012, instead, would lead the sum of the three masses to slightly exceed the mass of the beryllium nucleus, so that the latter would be stable. The same argument would be tackled again in several circumstances by the Joliot-Curies, and also by others (Irène Curie and Frédéric Joliot, "La complexité du proton et la masse du neutron", *CR*, *197* (1933), 237-238; Irène Curie and Frédéric Joliot "I. Production

The debate on the compound proton model involved – in an incidental way, for the time being – one of the protagonists of the neutron physics to come, through a note by Gleb Wataghin to the *Accademia dei lincei*. On April 23, 1933 A. Pochettino communicated to the *Accademia* a theoretical work by Wataghin, tackling, with relation to the β-decay, the compound proton suggested by Anderson. Wataghin considered β-decay to consist in a negative-positive electron pair creation, in the ejection of the negative electron, and in the capture of the positron by a neutron, which in turn changed into a (compound) proton. At the end of his note Wataghin wrote:[64]

> To sum up we think that, if the concept of intranuclear electrons […] has significance, it is more plausibly a question of the positive electrons, rather of the negative ones, also in the view that positive electrons are not ruled out to have null-spin and therefore to obey Bose statistics, in such a way that […] the neutron and the proton having spin 1 [in units h/4π] would be accounted for. *I owe the latter remark to Professor Fermi.*[65]

This is evidence of Fermi's interest in topics concerning the nucleus, and in particular β-decay, dating back to many months before he published the β-decay theory.[66]

Up to now we have dealt with some topics involving the concept of symmetry, such as the specular role played by positive and negative electrons, and the

---

artificielle d'éléments radioactifs. *II*. Preuve chimique de la transmutation des éléments", *JP*, 5 (1934), 153-156.

[64] Gleb Wataghin, "Sulla teoria del nucleo", *Rendiconti dell'Accademia dei Lincei*, 17 (1933), 645-647, on 646-647. Note that Wataghin's work was communicated some seven months before Leo Nedelsky and J. Robert Oppenheimer reported on the probability of internal pair formation (Leo Nedelsky and J. Robert Oppenheimer, "The production of positives by nuclear gamma-rays", *PhR*, 44 (1933), 948-949.

[65] Italics are added.

[66] Fermi, in his paper introducing weak interactions in physics, explicitly stated that the electron creation process «is in no respect analogous to the possibility that an electron-positron pair be created»; such a statement implicitly brings back to Wataghin's model. Enrico Fermi, "Tentativo di una teoria dei raggi β", *Nuovo cimento*, 2 (1934), 1-19, on 2.

symmetries, strictly relater to the latter, concerning the neutrons and the protons, which were ejected by light elements irradiated with α-particles. We are dealing now with one more case, relating to one further relationship in artificial disintegration processes involving neutrons. Such relationship was acknowledged at the time, and consists in the concept of reversibility of nuclear reactions.

In 1930, while Chadwick, J.E.R. Constable e E.C. Pollard were investigating proton emission by light elements irradiated with polonium α-particles, they came to the conclusion[67] that boron disintegration was due to the isotope $B^{10}$. In fact, energy evaluation led them to conclude that, if $B^{11}$ was disintegrated by α's, then the $C^{14}$ nucleus thus formed would have had a mass defect grater than $N^{14}$. Therefore, $N^{14}$ would have changed into $C^{14}$ by electronic capture, releasing energy. On the contrary, $N^{14}$ was abundant in nature, while $C^{14}$ completely unknown at the time.

Thus, once proton had been established to be ejected according to the reaction

$$He^4 + B^{10} \rightarrow C^{13} + H^1 \qquad [5]$$

Irène Curie and Frédéric Joliot, having discovered neutron emission from boron, only guessed that $B^{11}$ ejected a neutron in a similar way as $B^{10}$ ejected a proton. They did not investigate further their conjecture.[68] Also Chadwick limited himself to concluding that it was only «probable» that the heavy isotope of boron emitted neutrons.[69]

Empirical indication, concerning the boron isotope which could emit neutrons, was poor; nevertheless, when Feather discovered that neutrons caused nitrogen nuclei to disintegrate, he readily grasped that the latter reaction was the reverse reaction with respect to $B^{11}$ disintegration by α-particles: «We have observed the nuclear reaction

$$He^4 + B^{11} \leftrightarrow N^{14} + n^1 \qquad [6]$$

---

[67] Chadwick, Constable, and Pollard (ref. 28).

[68] Curie and Joliot (ref. 13).

[69] Chadwick (ref. 41), 701.

to take place both in the forward and reverse directions».[70]

It is worth noting that the reversibility principle in nuclear reactions then became foundational to a grater degree than the very experimental proofs which had led to propose it. On phenomenological grounds, uncertainty still survived on the reaction $He^4 + B^{11} \rightarrow N^{14} + n^1$, and then on the actual reversibility of [6]. Notwithstanding this Feather, when dealing with the transmutation of the carbon by neutrons, ascribed it to the light carbon isotope $C^{12}$, in accordance with such reversibility principle.[71] In fact, Feather explained that the reverse reaction with respect to beryllium transmutation by α-particles could actually take place, pointing out that the reversibility should hold in a similar way to what happened with reaction [6]:

$$He^4 + Be^9 \leftrightarrow C^{12} + n^1 \qquad [7]$$

In mid 1932 John D. Cockcroft and Ernest T.S. Walton dealt with another example of reversibility, even if in a broad sense: they succeeded in obtaining transmutation of light elements by accelerated protons, of energies of some hundred thousands electronvolts.[72] Cockcroft and Walton, having investigated lithium and beryllium transmutation through scintillation produced on a zinc sulphide screen by the ejected α-particles, and having noticed a weak effect also from fluorine and other elements, so commented: «The present experiments show that under the bombardment of protons, α-particles are emitted from many elements; the

---

[70] Feather (ref. 22), 725.

[71] Feather (ref. 38), 705-706. Really, the transmutation of carbon nuclei had been already observed by Harkins, Gans and Newson, although they did not state any reason why they had ascribed it to the isotope 12 according to the reaction $n^1 + C^{12} \rightarrow Be^9 + He^4$ (William D. Harkins, David M. Gans, and Henry W. Newson, "The disintegration of the nuclei of nitrogen and other light atoms by neutrons. I", *PhR*, *44* (1933), 529-537).

[72] John D. Cockcroft, and Ernest T.S. Walton, "Experiments with high velocity positive ions. II. – The disintegration of elements by high velocity protons", *PRS*, *137* (1932), 229-242.

disintegration process is thus in a sense the reverse process to the α-particle transformation».[73]

The reactions of disintegration by protons and α-particle emission, observed by Cockcroft and Walton, were the reverse – in a broad sense – of the first reactions ever investigated, consisting in capture of an α-particle and emission of a proton (though the reverse of $H^1 + Li^7 \rightarrow 2He^4$, of $H^1 + B^{11} \rightarrow 3\ He^4$ (or $Be^8 + He^4$), or of $H^1 + F^{19} \rightarrow O^{16} + He^4$ were not known, actually). Other important nuclear reactions, produced by accelerated protons, were investigated in the United States, and the concept of symmetry in 'artificial' nuclear reactions helps throw new light upon a deep connection of those results with the experiments to come on radioactivity induced by neutrons.

4. FINDING NEW ROUTES

In the United States investigations into neutrons, leading to remarkable results, were carried out in 1933, one year after the neutron had been discovered. In order to examine closely neutron scattering, the behaviour of neutrons passing through light substances (such as paraffin or water) and heavy ones (like lead) was investigated. Nuclear disintegration produced by neutrons was also examined, and the line of research which had been opened with particle accelerators by Cockcroft and Walton in Europe was extended to deuton beams, generating strong neutron sources.

Before we deal with the main experimental results attained on such topics in the United States, we shall examine a paper by Alfred Landé,[74] the renowned scholar of Arnold Sommerfeld, who had moved to the States a couple of years before. Interesting remarks of theoretical nature are formulated in that paper, which closely resemble in 'sensitivity' that kind of approach that, by driving questions into their essential lines, and drawing the most intuitive picture of physical phenomena, was so typical of Enrico Fermi as well.

The discovery of the neutron had fostered a deep renewal of nuclear models, independently of all the debates about the inmost nature of the neutron we have already recalled. In his paper, dated January 21, 1933, Landé tackled the constitution of nuclei, which he no longer considered to consist of the highest possible number of α-particles, zero up to possibly three protons, and electrons, as it was in models used before the discovery of the neutron; Landé, instead, now considered the nucleus as to consist of α-particles, neutrons, and zero or at most one proton. He closely examined the case of odd charged nuclei, tackling the question of their stability. Landé stressed that elements having a loose proton in the nucleus, *i.e.* elements with odd charge number, are much rarer than elements without a loose proton (even charge number); moreover, he stressed that the isotopes of even elements have (with due exceptions) an even number of neutrons. According to Landé, energy arguments alone were not enough to account for such an empirical rule: in fact, by relating the mass defects to the new nuclear model he had just examined, he noticed that adding neutrons to the nucleus caused the mass defect to increase linearly, of a quantity which could by quantified as 0.009 unit mass per neutron added, for many elements: «We have then to take for granted the preference of even arrangements of neutrons without an explanation by energy balances».[75] Another important aspect is tackled by Landé, relating to β-decay:[76]

---

[73] Ibid., 229.

[74] Alfred Landé, "Neutrons in the nucleus. I", *PhR*, *43* (1933), 620-623.

> If processes of **b**-*emission* connected with a transformation of one of the neutrons into a proton are taken into account, then the original loose proton and the new proton should associate with two neutrons to form an α-particle, and the odd element should pass over into an even one. However, though the final product has one α-particle more, it has three neutrons less. Now […] the energy gain 0.032 of building up an α-particle is about the energy loss of separating three neutrons and one β-particle from the nucleus; so these two effects may approximately balance each other, and we expect that minor secondary causes shall give the *decision* weather an element of odd charge number is or is not stable.

It is really remarkable that Landé accounted for β-decay claiming, as early as the beginning of 1933, a «transformation of one of the neutrons into a proton»: such a picture appears to anticipate, indeed, the concept that would become the foundational trait, together with the creation of an electron and a neutrino, of the theory that Fermi was to publish almost one year later. Apart from this, we can add that Landé, as he went by the mass defects, judged it to be a sort of accident, depending on «secondary causes», whether nuclei with odd atomic number (and with enough neutrons) either existed in nature, or were unstable.

---

[75] Ibid., 622.

[76] Ibid.

Concerning elements with odd charge number, the only ones known to have also an odd number of neutrons were $H^2$, $Li^6$, $B^{10}$, and $N^{14}$. Landé pointed out that[77]

> if there is only a small preference of even neutron arrangement before odd ones, this may give the decision mentioned above that β-emission in this case always will happen, hence odd arrangement of neutrons in odd elements will be unstable.

The question of the stability with respect to β-emission from the four stable isotopes recalled above was not a problem: according to Landé's nuclear model, such isotopes had only one loose neutron and, if they had decayed, the nucleus would have attained two loose protons, too many indeed. «Thus from the point of view of the neutron scheme the four exceptions prove the rule».[78]

To sum up, according to Landé a hypothetical odd charged element with an odd number of neutrons would have decayed through β-emission. As we shall see, this point of view would be reflected in Fermi's works on artificial radioactivity, but, still before that, in a paper reporting on fluorine disintegration by neutrons.

Having expounded the notable theoretical remarks by Landé, we are now dealing with some of the numerous experimental results achieved in the United States. We have already mentioned that many experiments were being concerned with the properties of neutron scattering.[79] J.R. Dunning e G.B. Pegram, for example, placed a lead cylinder near a small ionisation chamber, as a screen for neutrons from a Rn + Be source, and performed scattering measurements resorting to annular ring scatterers, coaxial with the lead cylinder: «As may be expected, paraffin and water, containing much hydrogen, show small backward scattering».[80] Dunning and Pegram paid much attention to the screen, which should just fill the geometrical path of the neutrons: absorption was largely scattering and every piece of matter outside the geometrical path would be able to deviate towards the chamber those neutrons directed elsewhere. Moreover, absorption curves revealed the probable existence of residual scattering from the room. In another work to come, Dunning recollected further results on neutron emission and scattering, and specified that «the scattering of the neutrons back into the ionization chamber or cloud chamber by adjacent matter, the walls of the chamber, nearby apparatus, walls, floors, etc. is usually important, and may often be as high as 10 percent or more».[81]

---

[77] Ibid.

[78] Ibid., 623.

[79] Neutron sources generally ranged from Rn + Be sources of many thousands of millicuries to Po + Be sources of few millicuries, depending upon the need as well on the availability. Some other kind of neutron sources, such as Po + $CaF_2$, was also attainable, in case properties of the neutrons from some particular element were to be investigated.

Another experimentalist, T.W. Bonner, examined how absorption of neutrons by paraffin, carbon, and lead depended on neutron velocity.[82] For that purpose, he resorted to neutron sources of three different kinds, namely Po + Be, Po + B, and Po + $CaF_2$, and to a ionisation chamber filled with different gasses. Bonner found paraffin to show the highest neutrons absorption. Neutron scattering cross section increased as neutron velocity decreased, both in paraffin and in carbon, but much more rapidly in paraffin. In lead, instead, high velocity neutrons showed larger absorption; Bonner explained that neutrons could lose only a small portion of their energy in elastic collisions with lead nuclei, and then that «the larger absorption in lead by neutrons of high velocity can be explained if we assume that the faster neutrons are more likely to make inelastic collisions than the slower ones».[83]

---

[80] John R. Dunning, and G.B. Pegram, "Scattering and absorption of neutrons", *PhR*, *43* (1933), 497-498, on 497.

[81] John R. Dunning, "The emission and scattering of neutrons", *PhR*, *45* (1934), 586-600, on 595-596. In this paper Dunning reported that, on account of geometry optimisation, he had put a tapered platinum 'cylinder' between the source and the chamber.

[82] T.W. Bonner, "Dependence of the absorption of neutrons on their velocity", *PhR*, *44* (1933), 235; T.W. Bonner, "Collisions of neutrons with atomic nuclei", *PhR*, *45* (1934), 601-607.

Such researches on neutron properties so far reviewed resemble those dealt with in Europe, strengthening the results already obtained there, however they did not yield fundamental novelties. More important contribution came from researches investigating nuclear disintegration processes, attained by resorting to both traditional neutron sources, and accelerated protons and deutons.

William D. Harkins, David M. Gans, and Henry W. Newson confirmed nitrogen disintegration to occur, stating:[84]

> On account of the ease of penetration of the neutron into the nucleus there seems to be no *a priori* reason, from this standpoint alone, to assume that low velocity neutrons should be ineffective but from the energy standpoint the higher velocities may be more effective.

In other words, neutrons are clearly stated here to have no hindrance in their penetration into the nucleus. Still, the difficulty persisted for them of letting the $\alpha$-particle ejected from the nitrogen nucleus, and the final boron nucleus gone in an excited state. More to say, «it is not improbable that some neutrons may attach themselves to atomic nuclei without causing a disintegration»:[85] Harkins, Gans and Newson were clearly referring to bare processes of neutron capture. Next, having reviewed reactions occurring with capture and without capture, they announced two more elements to be transmuted by neutrons, that is carbon and neon, according to the reactions:

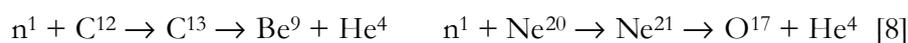

$n^1 + C^{12} \rightarrow C^{13} \rightarrow Be^9 + He^4 \qquad n^1 + Ne^{20} \rightarrow Ne^{21} \rightarrow O^{17} + He^4$ [8]

Elements disintegrated by neutrons now increased to four, since Feather had already proved nitrogen and oxygen to be disintegrated. It is worth stressing that neon was disintegrated by neutrons, although its nucleus should have consisted of five α-particles and then have been very stable.

---

[83] Ibid., 607.

[84] William D. Harkins, David M. Gans, and Henry W. Newson, "The disintegration of the nuclei of nitrogen and other light atoms by neutrons. I", *PhR*, *44* (1933), pp. 529-37, on 534.

[85] Ibid.

Results being attained on neutron physics were discussed at the 186[th] regular meeting of the American Physical Society, held in Chicago in June 1933 and characterised, on *Physical Review*, as «perhaps the most important scientific session in its history to date».[86] Prominent personalities, both from the United States and from Europe, attended the meeting: among them, we recall Aston, Bohr, Cockcroft, Compton, Harkins, Lawrence, Millikan, and Slater; from Italy, Tullio-Levi Civita and Enrico Fermi came, the latter discussing his theory of hyperfine structure.[87] Fermi, having attended the meeting, got certainly acquainted with results attained by those physicists who were engaged in frontier researches on nuclear physics; moreover, he was possibly stimulated and inspired for what concerned his experimental research to come in Rome.

Some of the talks are particularly noticeable, dealing with artificial disintegration of light elements by accelerated protons and deutons. Henderson, Lawrence, Lewis, and Livingston communicated, in three distinct accounts,[88] that they had observed evidence of disintegration in many substances: high speed protons caused α-particles ejection from lithium, beryllium, calcium fluoride, sodium chloride, aluminium, magnesium, and ammonium nitrate; accelerated deutons caused α-emission from lithium fluoride, ammonium nitrate, aluminium, beryllium, magnesium, sodium, chlorine, calcium fluoride, and boric oxide. In case

---

[86] "Proceedings of the American Physical Society. Minutes of the Chicago meeting, June 19-24, 1933", in *PhR*, *44* (1933), 313-331.

[87] See also: Enrico Fermi and Emilio Segrè, "Zur Theorie der Hyperfinstrukturen", *ZP*, *82* (1933), 728-749; Enrico Fermi and Emilio Segrè, "Sulla teoria delle strutture iperfini", *Memorie dell'Accademia d'Italia*, *4* (1933), 131-158.

[88] M. Stanley Livingston, Malcom C. Henderson, and Ernest O. Lawrence, "The disintegration of the elements by swiftly moving protons", *PhR*, *44* (1933), 316; Ernest O. Lawrence, and M. Stanley Livingston, "Three types of nuclear disintegration of calcium fluoride by bombarding protons of very great energy", *PhR*, *44* (1933), 316-317; Gilbert N. Lewis, M. Stanley Livingston, and Ernest O. Lawrence, "The disintegration of nuclei by swiftly moving ions of the heavy isotope of hydrogen", *PhR*, *44* (1933), 317.

of proton bombardment, it was warned that the specimens might contain some impurities.

The concept of 'symmetry', expounded in the previous paragraph, arouses our interest in reactions produced by protons and deutons. Let us consider, for example, the case of fluorine: in disintegration processes produced by $\alpha$-particles, two channels were involved, that is proton emission, and neutron emission – we indicate them as ($\alpha$, p) and ($\alpha$, n), respectively –; Cockcroft and Walton had shown the reaction (p, $\alpha$), with proton absorption and $\alpha$-emission, to take also place. Thus, in order to fulfil symmetry requirements, (n, $\alpha$) processes, consisting in neutron absorption and $\alpha$-emission, were still missing. More to say, ($\alpha$, p) and (p, $\alpha$) reactions transmuted fluorine into $Ne^{22}$ and into $O^{16}$ respectively, which were both stable nuclides; according to the Joliot-Curies, the reaction ($\alpha$, n) would end with the production of $Na^{22}$, not known in nature, and with a positive electron emission. It should be pointed out that the (n, $\alpha$) reaction would lead to the production of $N^{16}$, not yet known as well, and that $N^{16}$ would be an odd charged isotope, with an odd number of neutrons: precisely that kind of isotopes considered by Landé with relation to stability against $\beta$-emission.

Let us now consider aluminium: it was known that ($\alpha$, p) reactions transmuted aluminium into $Si^{30}$, and that ($\alpha$, n) reactions transmuted it into the previously unknown $P^{30}$, which in turn emitted a positive electron; now, (p, $\alpha$) reactions were discovered, which led to the production of $Mg^{24}$. The reaction (n, $\alpha$) was still missing: it would lead to $Na^{24}$, isotope with an odd atomic number, and an even mass number.

Sodium as well was known to suffer transmutation processes induced by $\alpha$-particles, both with proton emission and neutron emission. Still, in case of proton bombardment, sodium chloride was investigated, and not sheer sodium, so that the provenance of ejected $\alpha$-particles had no certain attribution.

It is plain that all transmutation processes concerning fluorine and aluminium, and involving neutrons as well – both in case of neutron capture, and in case of neutron ejection –, led, or would led, to isotopes not known in nature.

The present paragraph began with the exposition of Landé's theoretical treatment of the nucleus: we stressed that Landé pointed to β-decay, in accounting for the empirical rule that odd charged nuclei with an odd number of neutrons did not exist in nature (there being very few exceptions to the rule). In ending the present paragraph, we are now reviewing experiments, which can be regarded as the first case in which Landé's remarks could be explicitly applied.

On November 10, 1933 Harkins, Gans, and Newson communicated that, having irradiated with neutrons a cloud chamber filled with helium and difluor-dichlor-methane, they had observed paired tracks, springing ahead from the same point in the gas. «The momentum and energy relations, together with other evidence, indicate that most, and probably all these are disintegrations of fluorine nuclei».[89] Harkins, Gans, and Newson were considering as a reaction:

$$n^1 + F^{19} \to F^{20} \to N^{16} + He^4 \qquad [9]$$

It was clearly the (n, α) reaction that, beyond the three (α, p), (α, n), and (p, α) already known, was still missing for fluorine. As Landé had pointed out almost one year before, such reaction would lead to $N^{16}$, unstable through β-emission.

Harkins, Gans, and Newson compare the yield of disintegration processes produced by neutrons, in fluorine and in other elements. They drew a list, in which elements entered in order of decreasing yield, that was: nitrogen and fluorine, oxygen, neon, carbon, and chlorine. It is worth noting that Harkins, Gans, and Newson included chlorine in the list: (n, α) reactions in nitrogen, oxygen, neon, and carbon led to known isotopes, whereas (n, α) reaction in chlorine would not lead to any isotope known in nature (even though there were two chlorine isotopes available, with mass number 35 and 37). In other words, maybe that chlorine was proved to be, even earlier than fluorine, the first element suffering transmutation leading to the isotopes, of even mass and odd atomic number, that Landé had hypothetically regarded as undergoing β-decay. Unfortunately, in my researches I did not find any other report on chlorine disintegration by neutrons, apart from this very fleeting mention.

---

[89] William D. Harkins, David M. Gans, and Henry W. Newson, "Disintegration of fluorine nuclei by neutrons and the probable formation of a new isotope of nitrogen ($N^{16}$)", *PhR*, *44* (1933), 945-946, on 945.

As for $N^{16}$ produced by fluorine disintegration, Harkins, Gans, and Newson added:[90]

> The possibility should be mentioned that an electron, the track of which cannot be seen in the photographs, may be emitted at the time of the disintegration, in which case $O^{16}$ would be formed. However the assumption that nitrogen 16 is what is formed is more in accord with what has been found in the disintegration of nitrogen, oxygen, and neon, by neutrons. If nitrogen 16 is unstable it may disintegrate subsequently, but this would not affect the calculations given in this paper.

Harkins, Gans, and Newson conjectured that nitrogen 16 could also disintegrate, emitting an electron. The other case tackled by them is that of both α-particle and negative electron promptly emitted, as fluorine was transmuted into oxygen 16 by neutrons: it should be pointed out that such further possibility in some way resembles the Joliot-Curies stating that fluorine, sodium, and aluminium, irradiated with α-particles, could promptly eject one neutron and one positive electron.

One more remarkable result was announced in January 3, 1934. Lawrence and Livingston bombarded light elements – among which aluminium – by 3 million volt deutons. They observed large numbers of heterogeneous protons and neutrons coming out, that they ascribed to the disintegration of deutons. Furthermore, they obtained evidence of groups of protons. Aluminium, for example, emitted a group of protons having a range of about 68 cm air equivalent.[91] «These might well be the result of the reaction of the deutons with Al nuclei wherein neutrons are added to the Al nuclei and protons are emitted». Such reaction would led to the isotope 28 of aluminium, which does not exist in nature, and which would be related somehow to the artificial radioactivity induced by α-particles discovered by the Joliot-Curies.

---

[90] Ibid., 946.

[91] Ernest O. Lawrence and M. Stanley Livingston, "The emission of protons and neutrons from various targets bombarded by three million volt deutons", *PhR*, *45* (1934), 220.

# 5. THE DISCOVERY OF ARTIFICIAL RADIOACTIVITY INDUCED BY α-PARTICLES, DEUTONS, AND PROTONS

As we have already recalled, in June 1933 the Joliot-Curies discovered that positive electrons were emitted when aluminium and boron were irradiated with α-particles.[92] Since it was already known that both protons and neutrons might also be ejected, the Joliot-Curies ended up by stating that, sometimes, one positive electron and one neutron could be emitted, in place of one single proton. The same held for boron of mass number 10. These remarks led the Joliot-Curies to adopt Anderson's model for the proton,[93] consisting in one neutron and one positive electron. Wataghin pointed out that from that point of view, β-decay might have been interpreted as a pair creation, ending with positron capture by a neutron, thus turning into a proton. He had discussed with Fermi about the paper in which he expounded such conjecture.[94]

In October 1933, most relevant nuclear physicists of the time – among which Bohr, Bothe, Chadwick, Marie Curie, de Broglie, Dirac, Fermi, Gamow, Heisenberg, Pauli, Rutherford, Schrödinger, just to recall some – attended the seventh Solvay conference, held in Bruxelles. The Joliot-Curies in their communication, devoted to «Penetrating radiation from atoms under the influence of α-rays», tackled the topic of transmutation electrons.[95] The discussion to come was very lively, and Francis Perrin noticed that «experiments in which positive electrons are created apparently raise some difficulties, concerning energy conservation. Such difficulties resemble those which are typical of natural β-rays spectra».[96] As for energy spectrum, Perrin stressed that the energy of positive electrons, ejected by α's from thin aluminium layers, spread over a wide interval, ranging from $0.5 \cdot 10^6$ to $2 \cdot 10^6$ eV; neutrons from aluminium, instead, had energies not higher than 1 MeV. Thus, at variance with a suggestion by the Joliot-Curies, Perrin explained that the energy spectrum of positive electrons could not be accounted for, by simply assuming that energy divided between the positron and neutron. Neither could negative electrons account for the spectrum of positive electrons: negative electrons from aluminium did not exceed $10^6$ eV, actually.

---

[92] Curie and Joliot (ref. 61).

[93] Anderson (ref. 62).

[94] Wataghin (ref. 64).

[95] Irène Curie and Frédéric Joliot, "Rayonnement pénétrant des atomes sous l'action des rayons α", *Structure et propriétés des noyaux atomiques*, proceedings of the seventh Solvay conference – Bruxelles, October 22-29 1933 (Paris, 1934), 120-202.

[96] Ibid., 178.

> Thus, I do not believe that, referring to the emission of these electrons, one could account for the energy differences existing among the positive electrons. It seems reasonable, then, to suppose that the process proposed by Joliot decomposes itself into two consecutive emission processes: neutron emission, followed by emission of a positive electron, with production of an intermediate and unstable nucleus ($^{30}_{15}$P, in the case of aluminium); in a word, such nucleus would show radioactivity by positive electrons, and it should not be a surprise if one finds in such case a continuous spectrum, as in case of β-rays emitted in natural radioactive phenomena.[97]

Wolfgang Pauli was very sceptical about Perrin's prophetic account. He said that «the conclusion that there is a strong analogy between the positron emission occurring in artificial disintegration, and the spontaneous emission of β-rays, can hardly appear certain to me».[98] Bohr, instead, seemed to be interested in Perrin's proposal:[99]

> The question of being aware if energy conservation holds, in processes in which aluminium is bombarded by α-particles, is of the greatest importance. *There is no doubt* that the observation that the positrons do not all have the same velocity is not, *in itself*, an argument against energy conservation, *since we do not yet know how positron emission takes place*. If positrons really escape from the inside of the nucleus, as Joliot supposes, the circumstance would closely resemble those of β decay.

We can add that the analogy guessed by Perrin, between natural β-decay and emission of transmutation positrons, led to a full symmetry as for constitution of nuclei. The existence of transmutation positrons, in fact, had cast doubts on the fact that natural β-decay could be a good argument for the neutron to be compound;

---

[97] Ibid., 179.

[98] Ibid., 180.

[99] Ibid.

Wataghin, for example, in the paper he had discussed with Fermi, supposed the *proton* to be compound, consisting of a neutron and a positron, so that a pair was produced in β-decay, the positron being captured by the neutron. We can now widen such argumentation, concluding that the delayed emission of transmutation positrons, resembling natural β-decay, was not a good argument anymore, for the proton to be compound. Thus, we come to the conclusion that, if neither the neutron nor the proton were in a position to be considered as compound particles any longer, that would defy all previous accounts for β-decay, Wataghin's one included.

On January 1934 Irène Curie and Frédéric Joliot proved Perrin's conjecture:[100] positron emission from aluminium persisted after the α-source was removed, and faded away according to an exponential law. In their experiments, the Joliot-Curies placed a thin layer of aluminium 1 mm apart from a polonium α-particle source, irradiating the layer for 10 minutes; next, they resorted to a Geiger-Müller counter, having an orifice closed by a leaf of aluminium 7/100 mm thick, and noticed activity decaying according to an exponential law, with a mean-life of 3m 15s. They observed an analogous effect in boron and magnesium, to which they ascribed mean-lives of 14m, and 2m 30s respectively. In all three cases, the 60 mCi polonium supply induced an activity producing initially about 150 impulses per minute in the counter. Electrons from aluminium and from boron proved to be positively charged:[101] «It is probably the same for the radiation from magnesium». Mean-lives did not depend on the energy of α-particles.

> These experiments show that a new kind of radioactivity exists, with emission of positive electrons. We believe that the emission process is the following:

---

[100] Irène Curie and Frédéric Joliot, "Un nouveau type de radioactivité", *CR*, *198* (1934), 254-256.

[101] Ibid., 255. The Joliot-Curies resorted to the «method of the trochoid». It consisted in placing the specimen to be investigated between the cylindrical pole pieces of an electromagnet, close to the rim. A central screen was placed between the pole pieces. The field was not uniform at the rim, so that corpuscles described recurved trajectories, drawing epicycloids. A counter was opposed to the source with respect to the central screen. Corpuscles reached the counter, in accordance with the sign of their charge, along clockwise or anticlockwise paths.

$$ {}^4_2\text{He} + {}^{27}_{13}\text{Al} = {}^{30}_{15}\text{P} + {}^1_0\text{n}. $$

The ${}^{30}_{15}\text{P}$ phosphorus isotope would be radioactive, with a period of 3m 15s, and would emit positive electrons according to the reaction

$$ {}^{30}_{15}\text{P} = {}^{30}_{14}\text{Si} + e^+. $$

An analogous reaction might be considered for boron and magnesium, the unstable nuclei being ${}^{13}_{7}\text{N}$ and ${}^{27}_{14}\text{Si}$. The isotopes ${}^{13}_{7}\text{N}$, ${}^{27}_{14}\text{Si}$, ${}^{30}_{15}\text{P}$ can exist only for a very short time, and this is the reason why we cannot observe them in nature.[102]

It should be noted that in the case of magnesium Irène Curie and Frédéric Joliot's account did not come up to the real phenomenon. *A posteriori*, in fact, $Si^{27}$ has a mean-life of about 6s, and not 2m 30s, so that the Joliot-Cures did not observe $Mg^{24}$ transmuting – by α-capture and neutron emission – into $Si^{27}$, which in turn would have decayed into $Al^{27}$ by *positive* β-decay. What they observed was, instead, the *negative* β-decay of $Al^{28}$ – produced from $Mg^{25}$ by α-capture and proton emission – which occurred according to the following reactions:

$$ \text{He}^4 + \text{Mg}^{25} \rightarrow \text{Al}^{28} + p^1 \qquad \text{Al}^{28} \rightarrow \text{Si}^{28} + e^- (+ \bar{\nu}) \qquad [10] $$

The mean-life of $Al^{28}$, in fact, is 3m 13s (half-life 2m 14s), in much better agreement than the mean-life of $Si^{27}$ with that measured by Joliot-Curies (on March 20, 1934 they themselves finally stated that the radioactivity induced in magnesium was from $Al^{28}$). It is noticeable that the Joliot-Curie, though testing the sign of the particles coming from irradiated aluminium and boron, did not investigate here the signs of the particles emitted by the radioactive process induced in magnesium.

Fermi as well would fall into an analogous misunderstanding, while announcing artificial radioactivity induced by neutrons. Thus, such episodes denote two aspects relating to pioneer investigation on artificial radioactivity. The first aspect is the apparent resort to an economy principle, limiting the number of conceptual schemes necessary to frame the observed phenomena. That is: the transmutation of aluminium, boron, and magnesium went together with the emission of charged corpuscles; aluminium was already known (as boron too, now) to emit positrons; therefore, corpuscle emission could well be framed in positron emission phenomena, for all three cases. The second aspect is that those papers were sent for printing with great promptness, which might sometimes turn into haste, thus denoting a climate of straightforward competition among laboratories.

---

[102] Curie and Joliot (ref. 100), 255-256.

Let us now take the announcement of artificial radioactivity up again, noticing that the Joliot-Curies added a remarkable conclusion:[103]

> Some kinds of lasting radioactivities, analogous to the ones we have observed, can no doubt exist in case of bombardment by other particles. The same radioactive atom could no doubt be created through different nuclear reactions. The nucleus $^{13}_{7}N$, for example, being radioactive according to our hypotheses, might be obtained under the action of a deuton upon the carbon, followed by the emission of a neutron.

Neutron production by deuton bombardment had already been observed by H.R. Crane, C.C. Lauritsen, and A. Soltan, on beryllium and lithium chloride.[104] Still, in those two cases $Be^9$ would change into $B^{10}$, and $Li^7$ would lead to the production of two $\alpha$-particles, no unstable nuclei resulting. The isotope $N^{13}$, instead, which would result from the reaction $H^2 + C^{12} \rightarrow N^{13} + n^1$, was radioactive, as the Joliot-Curies had just proved.

Before proof or disproof came, of the new reaction that the Joliot-Curies had conjectured leading to $N^{13}$, they separated chemically the radioelements they had discovered.[105] Besides substantially confirming the three mean-lives they had already measured, they stated again, mistakenly, that radioactivity produced in the magnesium specimen should be exclusively ascribed to $Si^{27}$. In fact, they mentioned the separation methods, showing actually that the chemical behaviour of the radioactive nuclides produced in boron was the same as those of nitrogen, while aluminium transmuted into nuclides behaving like phosphorus; still, as for magnesium, the Joliot-Curies did not provide any precise account, generically stating that «analogous investigations are in progress with Mg».[106] In short, there were «both physical and chemical reasons pointing out that the nuclei $^{13}_{7}N$, $^{30}_{15}P$, and probably $^{27}_{14}Si$ should be radioactive, with positive electrons emission».

---

[103] Ibid., 256.

[104] H.R. Crane, C.C. Lauritsen, and A. Soltan, "Production of neutrons by high speed deutons", *PhR*, *44* (1933), 692-693.

[105] Irène Curie and Frédéric Joliot, "Séparation chimique des nouveaux radioéléments émetteurs d'électrons positifs", *CR*, *198* (1934), 559-561.

Irène Curie and Frédéric Joliot collected their results in a letter to *Nature*, dated February 10, 1934. They were still more precise, in foreseeing the development of researches concerning induced radioactivity to come: according to them, radioactive elements might also be produced «in different nuclear reactions with other bombarding particles: protons, deutrons, neutrons».[107] Thus, it was immediately guessed that not only α-particles would induce radioactivity.

Remarkable results concerning artificial radioactivity were attained also in the United States. Two papers, both dated February 27, 1934, reviewed the results attained in two Californian laboratories. The first of those two papers was by Malcom C. Henderson, M. Stanley Livingston, and Ernest O. Lawrence, from Berkeley University, and concerned «Artificial radioactivity produced by deuton bombardment».[108] Attempts to induce radioactivity by means of one and one-half million volts protons failed with most elements, except possibly in the case of carbon. Nevertheless, many substances were observed to emit both gamma-radiation and ionising particles for some time after having been irradiated by three million volt deutons: «In the light of the Curie-Joliot experiments these particles are presumably positrons». Measured half-lives were reported, though impurities in the specimens could not be excluded.

Concerning the substances that Henderson, Livingston and Lawrence had investigated, believing to have attained positive electrons emission from them, we are now dealing with the sole calcium fluoride, aluminium, and magnesium. As for calcium fluoride, they ascribed to the radioactive product a half-life of 40s, a value that, in spite of possible contamination of the specimen and of the very short period (which could cause trouble), was of the same order as the half-life of negative β-emission actually occurring, when fluorine undergoes radiative neutron-capture (such half-life is 11s). As for aluminium, they found that the half-life of the radioactive product was three minutes; now we know, *a posteriori*, that $Al^{28}$ produced by neutron capture from $Al^{27}$ undergoes negative β-emission with a half-life of 2m 18s; as for magnesium, lastly, Henderson, Livingston and Lawrence measured a half-life of 9m, whereas 9m 27s is the half-life of $Mg^{27}$, produced from $Mg^{26}$ by neutron capture, and emitting a negative electron.

---

[106] Ibid., 561.

[107] Irène Curie and Frédéric Joliot, "Artificial production of a new kind of radio-element", *Nature*, 133 (1934), 201-202.

[108] Malcom C. Henderson, M. Stanley Livingston, and Ernest O. Lawrence, "Artificial radioactivity produced by deuton bombardment, *PhR*, 45 (1934), 428-429.

To sum up, Henderson, Livingston and Lawrence believed they had observed positive electrons emission induced in calcium fluoride, aluminium, and magnesium. That would have been in agreement with the process foreseen by the Joliot-Curies for $C^{12}$, transmuting into the $β^+$-radioactive $N^{13}$ by deuton bombardment and neutron ejection. Nevertheless, *a posteriori* we come to the conclusion that in Henderson, Livingston and Lawrence's experiments $β^−$-radioactivity induced by neutrons was attained with aluminium and magnesium (and possibly with fluorine as well); indeed, activation occurred with accelerated deutons but not with accelerated protons (still of different energies). It is remarkable, in this respect, that the half-lives measured by Henderson, Livingston and Lawrence are almost the same as those that Fermi and collaborators would measure systematically, after they discovered slow neutrons and observed neutron radiative capture.

Again on February 27, Lauritsen, Crane and W.W. Harper, three experimentalists of the California Institute of Technology, reported on the activation of boron and carbon bombarded with 900 keV deutons.[109] The half-lives measured by Geiger-Müller were 20m and 10m respectively, and the tracks produced in a Wilson chamber revealed that positrons were emitted. Only two days later, Crane and Lauritsen wrote a letter to *Physical review*, describing the phenomenon in more details.[110]

---

[109] C.C. Lauritsen, H.R. Crane, and W.W. Harper, "Artificial production of radioactive substances", *Science*, *79* (1934), 234-235.

[110] H.R. Crane, C.C. Lauritsen, "Radioactivity from carbon and boron oxide bombarded with deutons and the conversion of positrons into radiation", *PhR*, *45* (1934), 430-432.

On March 14, 1934, Crane and Lauritsen reported, in another letter, they had succeeded in obtaining the activation of boron and carbon, by resorting to 900 keV protons as well.[111] They reasoned that the activation was not due to proton capture by $B^{10}$ and $C^{12}$; instead, $B^{11}$ and $C^{13}$ would capture a proton and eject a neutron. In fact: «One hesitates to accept the first type of proposed reaction, mainly because the probability of a particle being added to a nucleus without the ejection of some other particle to carry away the excess energy is extremely small». If Crane and Lauritsen had only irradiated, for example, aluminium with protons, they would have noticed that there was no way of obtaining the three minutes half-life measured by Henderson, Livingston and Lawrence on aluminium bombarded with deutons: *a posteriori*, we can infer that Crane and Lauritsen halted not far from concluding that neutrons could induce radioactivity in aluminium.

Artificial radioactivity with negative electrons emission was observed shortly after. In a paper received by the *Journal de physique* on March 20, 1934, even before Fermi's announcement of neutron-induced $\beta^-$-radioactivity, Irène Curie and Frédéric Joliot eventually reported that $\alpha$-particles irradiation led to the production of $Al^{28}$, decaying with emission of negative electrons.[112] Still on February 10, the Joliot-Curies had stated that the radioactivity induced in magnesium was due to $Mg^{24}$ absorbing an $\alpha$-particle, ejecting a neutron and thus transmuting into $Si^{27}$, in turn undergoing positive $\beta$-emission. Now, though not excluding the previous process, they suggested that $Mg^{25}$ instead absorbed an $\alpha$-particle, ejected a proton and transmuted into $Al^{28}$, which underwent negative $\beta$-decay. We come to the conclusion that the isotope 28 of aluminium had been obtained in two different ways, just by $\alpha$-irradiation and by deuton bombardment, but the former process was realised only on March by the Joliot-Curies, who thus announced artificial radioactivity with negative-electrons emission.

The discovery of artificial radioactivity was a fundamental achievement for nuclear physics, after the discovery of the neutron. Induced radioactivity was produced by the Joliot-Curies, by resorting to α-particles, by Henderson, Livingston and Lawrence by resorting to deutons, and by Crane and Lauritsen, by resorting to protons. Only one kind of reactions leading to artificial radioactivity was still missing, among those foreseen by the Joliot-Curies at the beginning of 1934:[113] «These elements and similar ones may possibly be formed in different nuclear reactions with other bombarding particles: protons, deutrons, neutrons».

---

[111] H.R. Crane, C.C. Lauritsen, "Further experiments with artificially produced radioactive substances", *PhR*, *45* (1934), 497-498.

[112] Curie and Joliot, "*I.* Production artificielle" (ref. 63).

6. THE PHYSICAL INSITUTE IN ROME GETS ACQUAINTED WITH THE DISCOVERY OF ARTIFICIAL RADIOACTIVITY

As we have already seen, in 1933 Enrico Fermi took part in both the meeting of the American physical society, and the seventh Solvay conference, held in June and in October respectively. Among the physicists attending in the Chicago meeting, Cockcroft, Harkins, Henderson, Lawrence, Lewis, and Livingston reviewed the researches in progress and the results already attained in their laboratories, being about to tackle radioactivity induced by accelerated ions within a few months. The topics discussed in Bruxelles regarded disintegration processes, produced by α-particles as well as by accelerated particles. Many questions of fundamental physics, ranging over the constitution of the neutron, as well as over general issues concerning the constitution of nuclei, were also reviewed; it was at this time that, for example, F. Perrin interpreted Joliot-Curies' results foreseeing the artificial radioactivity produced by α-particles.

When the discovery of artificial radioactivity was reported in Rome, the renowned physics institute located in via Panisperna – a picturesque street close to the centre of the city – was ready to welcome the announcement, and start a new course of researches. In fact, the Institute had lacked adequate supplies until short before, but thanks largely to Rasetti's efforts an adjustment of the equipment and supplies had been undertaken, thus making strong radioactive sources, Geiger-Müller counters and Wilson chambers available. Fermi's new course of researches in experimental nuclear physics could start.

The foundation «Oscar D'Agostino» in Avellino, Italy, keeps a really remarkable letter, that Fermi and Rasetti wrote to Oscar D'Agostino – the chemist of the group – on February 9, 1934.[114] D'Agostino, exhorted by Fermi and Rasetti, had left Rome for Paris in January, in order to get practice in radiochemical techniques at the *Institut du radium*. On February 9, Fermi and Rasetti reported to D'Agostino on progress at the physics institute in Rome, giving news that a Wilson chamber was ready, and the construction of another one was to be completed soon. Very remarkably, Fermi and Rasetti moreover wrote:

> We are preparing some counters, in order to repeat Joliot's experiments on artificial radioactivity with positron emission, and to inquire whether it is possible to separate the unstable radioactive product which should be produced, within the few minutes of its mean-life.

---

[113] Curie and Joliot (ref. 107), 202.

According to the letter to D'Agostino, therefore, as the artificial radioactivity was announced in Rome, Fermi and Rasetti did not promptly react by undertaking novel researches on neutron-induced radioactivity! Rather, they only planned, at first, to repeat Joliot's experiments on $\alpha$-induced radioactivity.

No other documentary proof, apart from this letter to D'Agostino, exists at present, of the fact that in February Fermi and Rasetti were to repeat the experiments carried out in Paris. This new version is particularly contrary, besides others, to an account by Emilio Segrè, according to whom Fermi, as soon as he got acquainted with the discovery of artificial radioactivity by the Joliot-Curies, «immediately saw that their work could be expanded tremendously by using neutrons as projectiles».[115]

---

[114] The Technical institute «D'Agostino» in Avellino, Italy, keeps several documents belonged to the chemist of Fermi's group, Oscar D'Agostino, who was native of Avellino. Noticeably, it was realised in 2002 that a notebook, along with a few spare sheets, largely consisted of laboratory notes by Fermi himself on the early experiments on neutron-induced radioactivity. As for the recovery of Fermi's notes, see: Giovanni Acocella, Francesco Guerra, and Nadia Robotti, "Enrico Fermi's discovery of neutron-induced artificial radioactivity: the recovery of his first laboratory notebook", *Physics in perspective*, 6 (2004), 29-41.

[115] See Emilio Segrè's introduction to Fermi's papers on neutron-induced radioactivity, in: Enrico Fermi, *Note e memorie (Collected papers)*, Edoardo Amaldi *et al.* eds. (2 vols., Roma-Chicago, 1962-1965), *1*, 639.

Rome, February 9, 1934

Dear D'Agostino,

we thank you for the news from you, and we are glad to hear that you could easily get your bearings in M.me Curie's Institute.

The procedure for preparing the bismuth plate is described in detail, in the note of which we are sending you two excerpts; in any case we are available for whatever further explanation. We are also sending you the photograph of a $\gamma$-rays spectrum, since reproduction is not very clear on Ricerca Scientifica. We are posting you, for the Joliots as well, a bismuth monocrystal, with one side already prepared for spectrography.

The polonium supply put in the Wilson chamber is very clean, giving a good fan of alpha particles, all of the same length; the large chamber works properly, but for the puttied parts. The other wire gauze-chamber of the Wilson type is at an advanced stage of construction. We are preparing some counters, in order to repeat Joliot's experiments on artificial radioactivity with positron emission, and to inquire whether it is possible to separate the unstable radioactive product which should be produced, within the few minutes of its mean-life.

Many wishes and greetings; please return our greetings to all our mutual acquaintances

*Enrico Fermi*

*F. Rasetti*

*Thanks for the postcards and many warmest greetings*

*E. Segrè*

*E. Amaldi*

The letter that Fermi and Rasetti wrote to D'Agostino on February 9, 1934, updating him about progress at the physical institute in Rome. At the bottom, Segrè and Amaldi's greetings. The note on *La ricerca scientifica* mentioned is: Enrico Fermi and Franco Rasetti, "Uno spettrografo per raggi «gamma» a cristalli di bismuto", *RiS*, 4 (1933), 299-302.

This letter draws attention to the need of integrating the personal recollections of those who took part in those events, occurred in Rome, with archive records. In this particular case, the letter makes one wonder what induced Fermi to redirect the research he was to start to verify $\alpha$-induced radioactivity, and why he turned his mind to investigate whether artificial radioactivity could be attained by resorting to neutrons instead. We might think that the very news coming from the United States, reporting on radioactivity induced by deutons and protons,[116] and perhaps also the induced $\beta^-$-radioactivity by the Joliot-Curies,[117] possibly triggered Fermi off, towards neutron-induced radioactivity with emission of negative electrons.

It is not reported elsewhere, that the experiments by Henderson, Livingston and Lawrence, those by Crane and Lauritsen, or even the latest by Joliot-Curies, could light the 'spark' making Fermi began to investigate whether neutrons could induce radioactivity. Fermi's intuition is rather related stright to his getting acquainted with the early Joliot-Curies' experiments on positive $\beta$-decay. However, the chronology of the events occurring at the beginning of 1934 seems to point out that some new significant clue should lead Fermi, who on February 9 appeared to be, together with Rasetti, still concerned with $\alpha$-particles, towards a different route: it appears quite reasonable that such significant clue consisted in the announcements coming from California, and possibly the latest from Paris.

One should bear in mind, however, that we cannot exclude the possibility that, on February 9, Fermi and Rasetti already had the intention of searching for neutron-induced radioactivity but that, fearing competition with the Joliot-Curies, simply preferred not to say a word of it with D'Agostino. In fact, one should remind that D'Agostino arrived at the *Insitut du Radium* in the same days when Irène Curies and Frédéric Joliot announced the discovery of artificial radioactivity. Thus, as D'Agostino himself would recall in 1958:[118] «It was not difficult for me to realise, from hundreds of very small pieces of evidence, that my coming had raised some perplexity. I could not understand why, at the time. I was not aware of what researches of the highest importance were being carried out […] on radioactivity phenomena».

---

[116] Henderson, Livingston, and Lawrence (ref. 108); Crane and Lauritsen (ref. 111).

[117] Curie and Joliot, "*I. Production artificielle*" (ref. 63).

[118] Oscar D'Agostino, "L'era atomica incominciò a Roma nel 1934 – 1ª puntata", *Candido*, *16*, n. 23, (June 8 1958), 20-25. «Non mi fu difficile capire da mille piccolissimi indizi che la mia venuta aveva destato qualche perplessità. Allora non seppi spiegarmi il perché.

We are now faced with two possibilities: either Fermi and Rasetti were not aware, still on February 9, of what kind of researches they were to begin with neutrons; or they had already planned to investigate neutron-induced radioactivity, but tactically didn't mention their true intentions in the letter to D'Agostino in Paris. We have no element for a definite decision between the two possibilities. Still, the first one seems more plausible, for reasons which will emerge in the course of the paper.

Once the announcement of artificial radioactivity induced by α-particles had reached Rome, it nourished the theoretical production at the physics institute in via Panisperna. At the beginning of March, Fermi communicated to the *Accademia dei lincei* a work, by Gian Carlo Wick, concerning an application of the newborn β-decay theory to the artificial radioactivity just discovered by the Joliot-Curies.[119] The circumstance that positron emission was concerned, and the continuous spectrum of the positrons coming from the $P^{30}$ produced from $Al^{27}$, both suggested that the theory recently published by Fermi was well suited to describe such new class of radioactive phenomena, induced by α-particles. Wick stressed that the theory, holding whether a neutron turned into a proton, and both an electron and a neutrino (name then used in place of «antineutrino») were created, held also for the reverse process, in which a proton turned into a neutron, and an electron and a neutrino were destroyed. Furthermore, for what concerned the radioelements discovered by the Joliot-Curies, Wick pointed out that the destruction of an electron of negative energy, and the resulting creation of a positive electron, «is not, as we shall see, the only way in which the nuclei in question may disintegrate».[120] Wick

---

Ignoravo l'importantissimo lavoro di ricerca che si stava conducendo […] attorno alla radioattività».

[119] Gian Carlo Wick, "Sugli elementi radioattivi di F. Joliot e I. Curie", *Rendiconti dell'Accademia dei Lincei*, 19 (1934), 319-324. I'm grateful to professor Nadia Robotti of the University of Genoa, for having drawn my attention to this paper.

[120] Ibid., 320

showed that decay processes induced by α's could not only consist in positron emission, but might also consist in the absorption of an orbital electron, and in the consequent emission of either an X-ray or a negative electron (by Auger effect). The mean lives of artificial radioactive nuclides could only roughly be evaluated; still, the theory allowed one to compare in a quite precise way the probabilities of such different decay modes to occur. Such is indeed the scope of Wick's work:[121]

> A doubt rises, whether the radioactive atoms produced be much higher in number than the emitted positrons actually counted (that may be of interest, in case one tries to evaluate the efficiency of such phenomenon, so as to make, for example, a comparison with the number of neutrons emitted in the disintegration process producing that radioactive element).

Wick's note shed due light on a trait that distinguished between artificial radioactivity and natural radioactivity. Being artificial radioactivity a secondary process, it was characterised by efficiency as a significant parameter; thus, the question reasonably arose, of evaluating to what extent the observed positrons be proper indicators of the number of the primary reactions really occurring. Wick calculated that a light nucleus, having captured an α-particle and having ejected a neutron, would have a probability of emitting a positron much higher than that of absorbing an orbital electron.

In the next paragraph, dealing with the experiments carried out in Rome, we shall go back to the concept of efficiency in inducing artificial radioactivity. Still, it is as from now plain that the same question reviewed in Wick's paper, communicated by Fermi before the *Accademia dei lincei* at the beginning of March, would arise also for the emission of negative decay electrons, induced by (n, α) reactions; such emission would be symmetrical to the emission of positive decay electrons, induced by (α, n) reactions.

Sources generally agree on stating that Fermi made up his mind in March, and tried to observe effects similar to those discovered in Paris, though resorting to

---

[121] Ibid.

neutrons. When Fermi took his Nobel lecture in 1938, he accounted for this start as follows:[122]

> [The Joliot-Curie] obtained the first cases of artificial radioactivity by bombarding boron, magnesium and aluminium with α-particles from a polonium source. […] Immediately after these discoveries, it appeared evident that α-particles very likely did not represent the only type of bombarding particles for producing artificial radioactivity. I decided therefore to investigate from this point of view the effects of the bombardment with neutrons. Compared with α-particles, the neutrons have the obvious drawback that available neutron sources emit only a comparatively small number of neutrons. […] This drawback is, however, compensated by the fact that neutrons […] can reach the nuclei of all atoms, without having to overcome the potential barrier, due to the Coulomb field. [Furthermore] their range is very long, and the probability of a nuclear collision is correspondingly larger than in the case of the α-particle or the proton bombardment. As a matter of fact, neutrons were already known to be an efficient agent for producing some nuclear disintegrations.

Such account, made some five years after those events, reflects the hallmark of the theoretician. Still, one can reasonably believe that the most cogent argument for neutron efficiency in producing transmutation was the last one, the most strictly experimental in nature, but precisely for this the most indisputable one: «As a matter of fact, neutrons were already known to be an efficient agent for producing some nuclear disintegrations». As in Feather's researches, many experimental results had already established that neutrons could produce (n, α) reactions in many elements: thus, it was «a matter of fact» that neutrons were so efficient, as to compensate (in part at least) for the disadvantage related to the weakness of neutron sources. As Irène Curie and Frédéric Joliot announced artificial radioactivity, «quite an obvious idea was that not only alpha particles could produce such artificial radioactivity, but probably also neutrons could».[123] Really, deutons and protons had already proved being efficient in inducing radioactivity; as for other more cases of artificial radioactivity, «only actual experiment could tell whether or not neutrons were good nuclear projectiles, and Enrico resolved to turn into an experimental physicist».[124]

---

[122] Enrico Fermi, "Artificial radioactivity produced by neutron bombardment", Nobel lecture; Fermi (ref. 115), 1038.

[123] Enrico Fermi, "Il neutrone", *Conferenze di fisica atomica (Fondazione Donegani) – Settima conferenza* (Roma, 1950), 90.

7. THE RADIOACTIVITY INDUCED BY NEUTRONS

Fermi announced the discovery of artificial radioactivity induced by neutrons in a short note to *La ricerca scientifica*, the scientific journal of the *Consiglio nazionale delle ricerche* (Cnr), entitled «Radioattività indotta da bombardamento di neutroni»:[125]

> I would like to report in this letter on some experiments, aimed to ascertain whether neutron bombardment produces phenomena of subsequent radioactivity, analogous to those the Joliots have observed with α-particles bombardment.

Fermi resorted to a neutron source, consisting of a glass bulb filled with beryllium powder and with 50 mCi of radium emanation (*i.e.* radon), the latter having been supplied by Giulio Cesare Trabacchi. Even though a very strong γ-radiation is emitted along with neutrons, such radiation «does not disturb in any way experiments like these», just because the phenomena searched for are subsequent to irradiation.

> Some small cylinders, which consist of the element to be investigated, are being exposed to radiations coming from that source, for a time varying from some minutes to some hours. Those cylinders are then rapidly placed around a wire counter. The counter has an outer wall consisting of a thin aluminium layer, about 0.2 mm thick, such that possible β-rays may go through and enter the counter. So far, the experiment has turned out well with two elements.

Fermi refers to aluminium and fluorine. As we have already stressed, fluorine and aluminium were the only two elements for which disintegration processes consisting in α-particle capture, followed either by neutron or proton ejection, as well as those consisting in proton capture followed by α-particle ejection, had already been proved to occur. In other words, fluorine and aluminium were the only two elements for which both (α, n), (α, p), and (p, α) reactions were already proved, the third one being the reverse of the second one. Moreover, fluorine and aluminium are both sheer elements, with one sole isotope each, and we can infer that that would greatly simplify the interpretation of experimental results. Fermi so continued:

> A small aluminium cylinder, irradiated with neutrons for a couple of hours, is then placed around the counter. During the early minutes, the cylinder causes a really considerable increase in the number of impulses, which increase by 30 or 40 per minute. Such effect fades away with time, reducing to the half in about 12 minutes. […] Having irradiated calcium fluoride for a few minutes, and having carried it close to the counter very quickly, it causes an increase in the number of impulses in the early moments. Such effect damps rapidly, reducing to the half in about 10 seconds.

---

[124] Laura Fermi, *Atoms in the family. My life with Enrico Fermi* (Chicago, 1954), 84.

[125] Enrico Fermi, "Radioattività indotta da bombardamento di neutroni", *RiS*, *5* (1934), 283.

Next, Fermi related induced radioactivity, he had just discovered, to a neutron capture followed by the ejection of an α-particle, *i.e.* Fermi's account conceived an (n, α) reaction to be causing induced radioactivity. In that way, Fermi conceived the production of isotopes still unknown, and having both odd charged nuclei and even mass number: exactly that sort of isotopes that, according to Landè's model,[126] were unstable, suffering negative β-decay. In fact:

> An interpretation of this phenomenon may be the following. Fluorine, having been bombarded with neutrons, disintegrates and ejects α-particles. The nuclear reaction is probably:
>
> 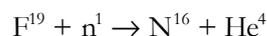
> $$F^{19} + n^1 \rightarrow N^{16} + He^4$$
>
> Nitrogen of weight 16 would be produced which, by a subsequent emission of a β-particle, might turn into $O^{16}$. An analogous account might hold for aluminium, in accordance with the possible nuclear reaction:
>
> 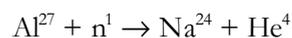
> $$Al^{27} + n^1 \rightarrow Na^{24} + He^4$$

---

[126] Landé (ref. 74).

The Na$^{24}$ thus produced would be a radioactive element, turning into Ca$^{24}$ through the emission of a β-particle. [...] Experiments are in progress, expanding such investigation to other elements, and examining the details of the phenomenon more closely.

It is plain that Fermi thoroughly adopted the current interpretative scheme, conceiving of disintegration processes by neutrons as consisting in the neutron absorption and α-particle ejection. More to say, neutron-induced radioactivity was attained with sheer nuclides of the $4n + 3$ type, of odd atomic number, and consisting of $n$ alpha particles, one proton and two neutrons. Actually, the reaction (n, α) conceived by Fermi would lead – or would have led – to the production of a new nuclide, having odd charge, once again, but one more loose proton. It is likewise plain, thus, that Fermi adopted one more interpretative scheme then current: in reviewing neutron-induced radioactivity, he implicitly adopted Landé's model, accounting for the transmutation of one loose proton and three neutrons into an α-particle. That is, as Landé had put it:[127]

> If processes of ***b**-emission* connected with a transformation of one of the neutrons into a proton are taken into account, then the original loose proton and the new proton should associate with two neutrons to form an α-particle, and the odd element should pass over into an even one.

Harkins, Gans, and Newson, on the other hand, had already obtained signs of the transmutation of F$^{19}$ into N$^{16}$, and had already conjectured that the latter would suffer a β-decay.[128]

Fermi, however, fell into a misunderstanding analogous to that in which Irène Curie and Frédéric Joliot had already fallen: the latter, in the case of radioactivity induced by α-particles in magnesium, inferred that they had observed the positive β-decay following an (α, n) reaction, whereas they had observed the negative β-decay following an (α, p) reaction; in a similar fashion Fermi mistakenly believed he had observed (n, α) reaction in aluminium. In fact, (n, α) reaction really occurs in Al$^{27}$, but we nowadays know that Na$^{24}$ so produced has a half-life of fifteen hours, whereas Fermi measured a half-life of about twelve minutes. Fermi had obtained a new kind of nuclear reaction, rather, consisting in a neutron absorption and the consequent emission of a proton, that is:

$$n^1 + Al^{27} \rightarrow Mg^{27} + p^1$$

---

[127] Landé (ref. 76).

[128] Harkins, Gans, and Newson (ref. 89).

The magnesium so produced has a half-life of nine minutes and a half, actually, not so different from that measured by Fermi.[129]

The driving of observative data into proven interpretative models could sometimes result in 'oversights' similar to those just reviewed. The latter might also result from the concern of communicating as soon as possible results just attained, though possibly still incomplete. Urgency likely struck Fermi, indeed: in order to complete the frame put forward by the Joliot-Curies when they discovered α-induced radioactivity, only neutron-induced radioactivity was missing; furthermore, many other researchers were engaged in the field of neutron physics and in the field of artificial radioactivity, abroad.[130]

One month and a half before Fermi's discovery, Fermi and Rasetti reported to D'Agostino on their plans concerning α-induced radioactivity. Urgency clashes with the possibility that, on February 9, they had already decided rather to investigate neutron-induced radioactivity, thus wasting various weeks before they actually started to. Rasetti, having been involved by Fermi in their first attempts to observe neutron-induced radioactivity,[131] should leave for Morocco not earlier than March 20,[132] while Fermi continued the experiments. Fermi's discovery should occur between March 20 and 25, while the first attempts should be carried out, reasonably, only a few days earlier. Thus, one might wonder why Fermi and Rasetti should wait until mid-March, if really they had planned such important experiments since February 9. For this reason, it is plausible that Fermi and Rasetti had not decided yet to search for neutron-induced radioactivity, on February 9 (when various counters were, or were to be, in operation), in accordance with what they actually wrote to D'Agostino.

---

[129] In a second note, Fermi stated that he had obtained radioactivity also in case of other elements, among which iron and phosphorus: in case of those two elements, chemical analysis carried out by D'Agostino related the decays to a neutron absorption, followed by the ejection of a proton, instead of the ejection of an α-particle. The first paper which was signed also by Fermi's collaborators was dated May 10; it was pointed out there that aluminium showed one more activity, with half-life of the order of one day, besides that of half-life of twelve minutes. On June 23, the long half-life activity of aluminium was ascribed to Na$^{24}$, whilst it was realised at last that the short activity that Fermi had measured in March was «probably» related to the production of Mg$^{27}$ (Enrico Fermi, "Radioattività provocata da bombardamento di neutroni", *RiS*, 5 (1934), 330-331; Edoardo Amaldi et al. "Radioattività «beta» provocata da bombardamento di neutroni – III", *RiS*, 5 (1934), 452-453; "Radioattività provocata da bombardamento di neutroni – IV", *RiS*, 5, 652-653.

[130] One more piece of evidence of haste comes from a misprint on the paper of March 25, where Fermi wrote that, following on the transmutation of Al$^{27}$, «the Na$^{24}$ thus produced would be a radioactive element, turning into Ca$^{24}$ through the emission of a β-particle». It is plain that he meant Mg$^{24}$ instead.

An interesting historical source came out in 2002, held by D'Agostino and, after his death, donated, along with other documents belonged to him, to a technical institute in Avellino, his native town. Such source consists of some laboratory records by Fermi, and in particular of a notebook, recording a great part of the experimental work carried out by Fermi during the first month of his researches on neutrons, including the measures proving that he had attained induced radioactivity.[133] For what concerns the discovery of neutron-induced radioactivity, as well as the preliminary phases of such discovery, it is then largely advisable to carry out a comparison of our previous account with the documents kept at the Foundation «Oscar D'Agostino», at the homonymous technical institute in Avellino.[134]

---

[131] Actually, Laura Capon Fermi and Emilio Segrè's early accounts in mid-1950s stated that Fermi tried to induce radioactivity on his own from the very beginning. Notwithstanding this, in 1962 Rasetti recalled that he and Fermi began the search for neutron-induced radioactivity together. All the accounts to come, Segrè's included, would be in agreement with Rasetti's version (Laura Fermi (ref. 124), 85; Emilio Segrè, "Fermi and the neutron physics", *Review of modern physics*, *27* (1955), 257-263, on 258; Rasetti, in Fermi (ref. 115), 549; Emilio Segrè, "The consequences of the discovery of the neutron", *Proceedings of the tenth international congress of the history of science* (Paris, 1962), 149-154, on 151; Edoardo Amaldi, "From the discovery of the neutron to the discovery of nuclear fission", Physics report, 111, (1984), 124).

[132] We can reconstruct quite easily from archive records the date when Rasetti's left Rome. In fact, the archives of the University «La Sapienza» in Rome keep Rasetti's record book relating to his course of spectroscopy for the academic year 1933-1934, reporting that his last lesson before Easter holiday was on March 20. Furthermore, it is generally stated that Rasetti left for Morocco just for a holiday, after his and Fermi's first unsuccessful attempts to produce neutron-induced radioactivity. Still, the Central archive of state in Rome keeps documents showing that he went to Morocco in his capacity as Italian delegate at the 58th meeting of the *Association française pour l'avancement des sciences*, held in Rabat from March 28 to 30 (I'm grateful to professor Gianni Battimelli, of the physics department of «La Sapienza», for this useful piece of information).

The first pages of the notebook record the attempts to obtain a proper set up of the amplifying circuits; the latter, including three vacuum triodes, connected the Geiger-Müller counters to telephonic counters (recording the impulses coming from the geigers). Some tests were also carried out, in order to evaluate the background radiation revealed by the geigers, under various shielding. Fermi, as he had attained a first set up of the devices, tackled the question of the sensitivity of the Geiger-Müller counters, which were expected to fulfil some minimum requirements: granted that neutron induced radioactivity was produced, it might well be very weak; thus, in order to undertake an effective search of it, one had to check preliminarily that the devices would work properly, revealing also very weak $\beta$-activities. In order to verify if counters were sensitive enough, it was then necessary to have a sample of a substance with a very weak $\beta$-activity, to be chosen as standard. The substance fixed on by Fermi was a solution of potassium-chloride.

Various measurements on the activity of KCl are recorded on Fermi's notebook, one being present also on page 18, where, as Acocella, Guerra, and Robotti have pointed out,[135] the first attempt – without result – of attaining the activation of a platinum sample is also reported. However, some measurements on aluminium which are recorded on the following page clearly show that in that case the irradiation with neutrons initially let the counts rise, from about ten per minute of the background, to about sixteen per minute, the effect fading away with time (fig. 1). Lead, calcium fluoride, mercurous chloride, and potassium chloride – the latter not irradiated, though with various shielding settlements – were investigated next. Among those substances, the only two which then showed to be activated were aluminium and calcium fluoride. Really, Fermi strove and examined them more closely. Aluminium, moreover, was irradiated with neutrons filtered through 1 mm of lead, as well as, during a whole night, with sole γ-rays, and in the latter case aluminium did not activate: Fermi wrote in his first paper that the strong γ-radiation from Rn + Be source «does not disturb in any way experiments like these» being founded on experimental grounds.

---

[133] Acocella, Guerra, and Robotti (ref. 114). The experiments on neutrons carried out by Fermi during the early weeks, however, are not completely recorded in the available documents: some minor records, like the measurements on selenium and antimony, are apparently missing for example, along with the records from the early, unfruitful attempts by Fermi and Rasetti.

[134] A first analysis of Fermi's notebook has been already carried out by Acocella, Guerra, and Robotti (ref. 114), and the interested reader can find there a detailed report on the measurements taken by Fermi on fluorine and aluminium.

[135] Acocella, Guerra, and Robotti (ref. 114), 36.

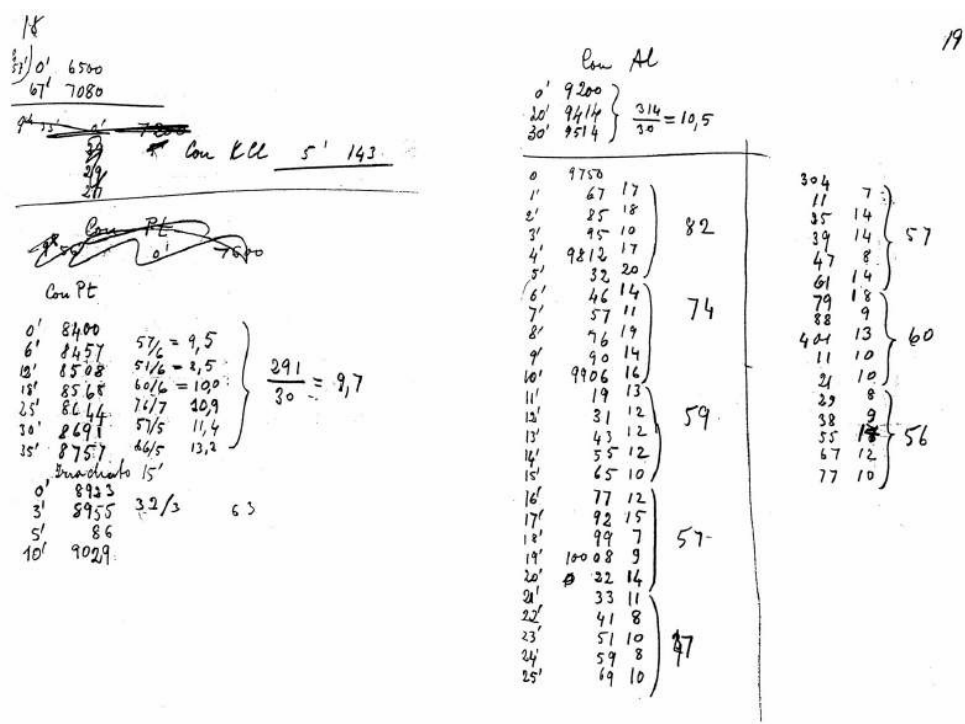

FIGURE 1. – On the left, page 18 of the notebook. The measurement on a solution of potassium chloride showed the sensitiveness of the counter. Then Fermi attempted to activate platinum. On the right, aluminium was activated (figure from Acocella, Guerra, and Robotti, (ref. 114)).

The first substances investigated by Fermi were, of course, all in the solid state, in order to avoid the use of containers (which in turn could possibly activate). One might wonder why Fermi chose to start investigating just platinum, aluminium, lead, calcium fluoride, and mercurous chloride. As for fluorine and aluminium, we have already reviewed some properties that might account for their use by Fermi, that is, fluorine and aluminium might suit (n, $\alpha$) reactions leading negative $\beta$-decay (besides being sheer elements). As for aluminium, moreover, we can recall that the walls of the Geiger-Müller counters were generally made of such metal, so that, if aluminium had not become radioactive, one would have gained a great advantage, being in a position to irradiate substances directly close to the counters (that would have been particularly advantageous, for example, in the search for activities with very short mean-lives). As for platinum, lead, and mercury, let us consider this: of each element, take the heaviest isotope (being not in negligible amount in nature), and then calculate the quantity $\delta = (A - 2Z)/Z$ (that is, consider the number of neutrons in excess, with respect to the hypothetical isotope having as many protons as neutrons, then calculate the ratio of such neutron excess to the atomic number). Now, among all the elements of the periodic table, those with the highest value for the relative excess of neutrons are, in order of decreasing $\delta$: mercury ($\delta = 0.550$), platinum ($\delta = 0.538$), and lead ($\delta = 0.536$). Let us consider that, also in the light of Landé's model, an excess of neutrons inside the nucleus might favour $\beta$-decay. It is then plain that, in order to attempt to induce artificial radioactivity (even in even elements), it was better to irradiate nuclei for which the relative excess of neutrons was already large: that is, just mercury, platinum, and lead. To sum up, we can point to very simple reasons, why Fermi chose to search neutron-induced radioactivity just in fluorine, aluminium, platinum, mercury, and lead at first. Simplicity, so characteristic a trait of Fermi's way of doing physics, here comes out once more.

By turning Fermi's notebook over (which would in fact be right side, actually), there are some fifteen pages focusing on the $\beta$-decay theory that Fermi had just published, between the end of 1933 and the beginning of 1934. The calculations on the notebook concerned the «probability of a transition to occur, in which the $\beta$-particle is captured by a K orbit» (fig. 2). On the following page Fermi reported an expression for the mean-life of light nuclei, which he evaluated as a function of the nuclear charge, and for various energies of the decay electron. Thus, the same notebook is reporting records from the experiments that revealed the existence of neutron-induced radioactivity, as well as calculations on the β-decay of light elements.

$$\text{Termine 1s relativistico}$$
$$\begin{cases} F = f\, r^s e^{-\lambda r} \\ G = r^s e^{-\lambda r} \end{cases} \qquad s = \sqrt{1-\gamma^2}-1 \ ; \ \gamma = Z/137$$
$$f = -\frac{s}{\gamma}$$
$$\lambda = \frac{mc^2 \gamma^2}{Ze^2}$$
$$\tilde{\psi}\psi = \frac{(2\lambda)^{3+2s}}{4\pi\, \Gamma(3+2s)}\, r^{2s} e^{-2\lambda r}$$

Probabilità di transizione in cui la particella β viene legata in una orbita K

$$P = \frac{8\pi^3 g^2}{h^4} \left| \int v_m^* u_m d\tau \right|^2 \frac{p_\sigma^2}{c}\, \tilde{\psi}_s \psi_s$$

Trattando l'elettrone K non relativisticamente e ammettendo che il neutrino emesso abbia energia quasi eguale risulta

$$P = \frac{2048\, \pi^{12} g^2 m^5 e^{14}}{c^3 h^{14}}\, Z^7 \left| \int v_m^* u_m d\tau \right|^2$$

$$P \sim 10^{-20}\, Z^7 \left|\ \ \right|^2$$

FIGURE 2. – The result of Fermi's calculations on the probability that a decay electron was absorbed into a K orbital (figure from Acocella, Guerra, and Robotti (ref. 114)).

The calculation of the probability for a decay-electron to be captured by a K-orbital can be accounted for in a very simple way. At first, let us recall Wick's work communicated to the *Accademia dei lincei* by Fermi on March 4, concerning the radioactive elements discovered by the Joliot-Curies: it reports the calculation of the probability for an orbital electron to suffer K-capture by an unstable nucleus.[136] The nuclides produced by the α's could either emit positrons, revealing the decay, or absorb orbital electrons. Now let us note that, in exactly the same way, neutron-induced radioactivity would lead either to decay-electrons leaving the nuclei, thus revealing the activity, or to electrons absorbed in K orbitals, thus unobservable.

---

[136] Gian Carlo Wick (ref. 119).

However, a fundamental difference existed between radioactivity induced by α-particles, discussed by Wick, and radioactivity induced by neutrons: positive β-decay had already been established to occur by the Joliot-Curies, while induced negative β-decay was still not established, when Fermi communicated Wick's work. Thus, as for the former phenomenon, one might wonder whether radioactive nuclei even exceeded emitted positrons; as for the latter phenomenon, the opposite question, much more severe, arose: even if one activated light nuclei by neutrons, could he observe negative β-emission indeed? Or the number of negative electrons leaving the atom would have been too small, in comparison with the unstable nuclei really produced, so that one would hardly have observed neutron-induced radioactivity?

When Fermi published his β-decay theory, he calculated the probability that an electron was emitted in an unboud state by heavy elements. Now he had «resolved to turn into an experimental physicist», but it is not given to us to know if he had already been acquainted with the discovery of negative β-decay, announced by the Joliot-Curies on the paper of March 20.[137] In any case, to be sure that there was no theoretical hindrance, for the neutron-induced radioactivity possibly produced to be eventually observed, he had to answer those questions just mentioned above, evaluating the probability that a decay-electron was absorbed into a K orbital: that is exactly the calculation he reported in the notebook.

Let us note that one could simply try and search experimentally for neutron-induced radioactivity, without fearing for pieces of hindrance in advance. Therefore, the calculation reported on the notebook supports the following view: having Fermi (and Rasetti, then left for Morocco) made some unsuccessful trials with a Po + Be source, Fermi investigated whether there existed some theoretical hindrance, instead of being the experimental supplies lacking. The calculation gave a very small probability for K-capture to occur, and that strongly encouraged Fermi, who eventually succeeded in observing neutron-induced radioactivity. Fermi was in a position to combine physics intuition and experimental ability with a mastery of β-decay phenomena, so that he managed to exploit his theoretical knowledge also in carrying out experiments.

---

[137] Curie and Joliot, "*I. Production artificielle*" (ref. 63).

CONCLUSIONS

In the present paper we have focused on dawning neutron physics and on the events leading to the discovery of artificial radioactivity phenomena. During the two years preceding the discovery of the latter, intense experimentation was carried out on neutrons. Researches concerned neutron absorption, scattering cross-section, and efficiency in producing nuclear transmutation. Some more features seem remarkable with relation to experiments to come, and to the discovery of neutron-induced radioactivity. One of those features concerns the many $\beta$-traces produced by the penetrating radiation from beryllium. From the very beginning physicists faced concern for electronic tracks related to beryllium radiation, observed not later than the end of 1931. Such tracks were at first merely related to a Compton effect produced by the $\gamma$-rays comprising beryllium radiation. Still, tracks apparently belonging to backward electrons were also discovered. Some ingenuous accounts were proposed, even though it was the discovery of the positron that settled things more clearly. By reviewing some of the results of the time, however, it is not excluded that visible electrons sometimes resulted from nuclear activation produced by neutrons.

Another interesting topic proper to nuclear physics and dawning neutron physics was the concept of 'reversibility' for nuclear reactions. The occurrence of a given reaction involving neutrons was claimed, in some cases, by merely pointing out that the reverse reaction was already known to take place. Remarkably, nuclei resulting from some of these reactions produced by neutrons fell within the domain of a clear treatment of nuclear stability, proposed by Landé.

To sum up, we dealt with some meaningful topics of the nuclear physics practised during the two years that preceded the discovery of neutron induced radioactivity, and which clearly represented a substratum for it. The appearance of electronic tracks as a general feature of experiments involving penetrating radiation from beryllium, the fact that the phenomenology of the reactions produced by neutrons was currently reviewed, and, what's more, the fact that a treatment existed, accounting for possible unstable nuclei produced in such reactions: they are all occurrences that may well have contributed to form the background of knowledge, that Fermi possessed when he discovered artificial radioactivity induced by neutrons. In this respect, the role of his participation in important congresses should not be disregarded either.

As for neutron-induced radioactivity, many interesting pieces of information about Fermi's work may be drawn from archive records. There seems to be evidence that Fermi, having planned to repeat the experiments on α-induced radioactivity, eventually switched to neutrons instead. The possibility is here reviewed, that Fermi changed his plans because for the time being – *i.e.* from the discovery of α-induced radioactivity to that of neutron-induced radioactivity – he got acquainted with the new results on artificial radioactivity obtained by means of accelerated deutons and protons (and perhaps, even with the discovery of artificial radioactivity with negative electrons emission). A review of experiments on deuton and proton-induced radioactivity is given, showing that in some cases half-lives were, *a posteriori*, very similar to the half-lives later obtained with neutrons. There is no definite proof that Fermi did eventually switch to neutrons, having considered α-induced radioactivity at first; so much the less that it were later results which let him turn to neutrons. Nevertheless, signs exist which strongly suggest that things might go in such a way.

Many laboratories were engaged in researches concerning artificial radioactivity and neutron properties, at the time. Electronic traces, transmutation produced by neutrons, matter of nuclear stability and artificial radioactivity, were all topics handled there; still, the discovery of neutron-induced radioactivity remarkably came from the laboratories of via Panisperna in Rome, which had just got ready to start researches in nuclear physics. The very fact that the adjustment of the equipment and supplies for nuclear purposes was concluded in Rome shortly after the discovery of artificial radioactivity should represent a very favourable circumstance for neutron-induced radioactivity to be discovered there. Furthermore Fermi, precisely because he was turning to some experimental researches in which he had not been engaged before, could freely direct all his efforts towards so a precise goal, as the search for neutron-induced radioactivity was. Really, Fermi went through searching for neutron-induced radioactivity even though his early attempts with Rasetti had been unfruitful. Versatility and simplicity were both characteristic of Fermi's way of doing physics. As for the former, it is stressed that it is possible that Fermi was deeply encouraged by his own theory of β-decay, in pursuing experiments which had revealed unfruitful at first. As for the latter, simplicity comes out in the choices he made in order to investigate neutron-induced radiaoctivity.

To sum up, we have gone through numerous works concerning neutron physics in the early 1930s, so that one may make out the imaginary lines joining them the one to the other: one may eventually notice how advancement is strictly related to results previously attained, preparing the path to, and rendering possible, new achievements. In other words, we devoted ourselves to drawing here an outline of how, «through precise and almost immediate indications arising from experience», the first evidence concerning dawning neutron physics appeared.